\documentclass{mn2e} 
\usepackage[dvips]{graphicx} 
\usepackage{amssymb} 
\usepackage{amsbsy} 
\usepackage{txfonts} 
\newcommand{\apj}{{ApJ}} 
\newcommand{\mnras}{{MNRAS}} 
 
\newcommand{\msun}{{\rm M}_{\sun}} 
\newcommand{\ledd}{L_{{\rm Edd}}} 
\newcommand{\medd}{{\dot{M}_{\rm Edd}}} 
\title[]{X-ray spectra of hot accretion flows}   
\author[]  
{
Andrzej Nied\'zwiecki,$^1$\thanks{E-mail: niedzwiecki@uni.lodz.pl (AN),
fgxie@shao.ac.cn (FGX), agastepnik82@gmail.com (AS)} 
Fu-Guo Xie$^{2}$\footnotemark[1]
and Agnieszka St\c epnik$^{1}$\footnotemark[1]\\  
$^1$Department of Astrophysics, University of \L \'od\'z, Pomorska 149/153, 
90-236 \L \'od\'z, Poland\\  
$^2$Key Laboratory for Research in Galaxies and Cosmology, Shanghai
  Astronomical Observatory, Chinese Academy of Sciences,\\  
80 Nandan Road, Shanghai 200030, China\\  
}

\date{6 June 2014} 
\pagerange{\pageref{firstpage}--\pageref{lastpage}} 
\pubyear{} 
\begin{document} 
\maketitle 
\label{firstpage} 

\begin{abstract}

We study radiative properties of hot accretion flows in a general relativistic model with an exact treatment of global Comptonization, developed in our recent works. We note a strong  dependence of electron temperature on the strength of magnetic field and we clarify that the underlying mechanism involves the change of the flow structure, with more strongly magnetised flows approaching the slab geometry more closely. We find that the model with thermal synchrotron radiation being the main source of seed photons agrees with the spectral index vs Eddington ratio relation observed in black hole transients below 1 per cent of the Eddington luminosity, $\ledd$, and models with a weak direct heating of electrons (small $\delta$) are more consistent with observations. Models with large $\delta$ predict slightly too soft spectra, furthermore, they strongly overpredict electron temperatures at $\sim 0.01 \ledd$. The low-luminosity spectra, at $\la 0.001 \ledd$,  deviate from a power-law shape in the soft X-ray range and we note that the first-scattering bump often resembles a thermal like component, with the temperature of a few hundred eV, superimposed on a power-law spectrum. The model with thermal Comptonization of thermal synchrotron radiation does not agree with well studied AGNs observed below  $\sim 0.01 \ledd$, for which there is a substantial evidence for the lack of an inner cold disc. This indicates that the model of hot flows powering AGNs should be revised, possibly by taking into account an additional (but internal to the flow) source of seed photons.
\end{abstract}

\section{Introduction} 
\label{intro} 

The nature of accretion flows powering black-hole systems changes around the bolometric luminosity of $\sim 10$ per cent of the Eddington value.
At lower luminosities, signatures of an optically thick disc disappear, or weaken significantly,  and then these lower-luminosity systems are supposed to be powered by optically thin, hot accretion flows (see, e.g., Narayan \& McClintock 2008, Yuan \& Narayan 2014; hereafter YN14). The origin of the X-ray radiation, typically dominating  the radiative output at low luminosities, remains a not fully understood issue, in spite of substantial theoretical work and observational advances during the last two decades. Its observed properties, in particular high-energy cut offs typical for a thermal electron plasma, favour thermal Comptonization in the inner parts of hot flows  (e.g.\ Zdziarski \& Gierli\'nski 2004). Alternatively, the X-rays may come from a jet (e.g.\ Markoff, Falcke  \& Fender 2001, Markoff, Nowak \& Wilms 2005); see e.g.\ Russell et al.\ (2010), Zdziarski et al.\ (2011), Gardner \& Done (2013), Veledina, Poutanen \& Vurm (2013) for related discussions.

The major problem for the hot flow model, pointed out e.g.\ by Yuan \& Zdziarski (2004, hereafter YZ04), concerns high values of the predicted electron temperatures, and the corresponding cut-off energies being much higher than what is observed. However, previous studies of this issue relied mostly on simplified models, in particular involving the use of a pseudo-Newtonian potential  as well as local approximations of Comptonization, both being potential sources of serious inaccuracies. 

In this work we systematically investigate the X-ray emission using our recently developed  model aimed at precise modelling of spectral formation in hot flows (Nied\'zwiecki, Xie \& Zdziarski 2012; hereafter N12, see also Xie et al.\ 2010). The model involves an exact, Monte Carlo (MC) treatment of global Comptonization with general relativistic (GR) transfer effects as well as global solutions  of the fully GR hydrodynamical model. We use a hot-flow model essentially following the original formulation of advection dominated accretion flow (ADAF) scenario by Narayan \& Yi (1994), and its later development e.g.\ by Narayan \& Yi (1995), Abramowicz et al.\ (1996), Esin, McClintock \& Narayan (1997), i.e.\ we consider a two-temperature flow with cooling of electrons determined by thermal synchrotron and bremsstrahlung emission and their thermal Comptonization. Even for such a standard version of the model,
the dependence on key parameters has not been fully studied in a precise model.

We consider the dependence on the poorly understood parameters in ADAF theory describing the strength of magnetic field and the fraction of the viscous dissipation that directly heats electrons. Concerning the plasma magnetization, some magnetohydrodynamic simulations (e.g.\ Hawley \& Krolik 2001) indicated a rather weak magnetic field, with a high ion to magnetic pressure ratio, $\beta$. Therefore, many recent investigations of ADAFs, including our previous works, are limited to such high-$\beta$ cases. On the other hand, a more general conclusion from numerical experiments is that the $\beta$ parameter is tightly correlated with the viscosity parameter, $\alpha$, namely $\alpha \beta \sim 0.5$ (see references and discussion in YN14). This, with the requirement of a high value of viscosity parameter in luminous hot-flows, $\alpha \ga 0.1$ (cf.\ YZ04), indicates that a value closer to equipartition, $\beta \sim 1$, may be more relevant. We thoroughly investigate the effect of magnetic field strength by comparing solutions for an equipartition and a weak (contributing 1/10th of the total pressure) magnetic field, we also consider a case of magnetically-dominated flow (cf.\ Oda et al.\ 2010).

Furthermore, we take into account outflows, which can be expected on both theoretical and observational grounds  and we assume the strength of an outflow supported by the polarization measurements in Sgr A* (see review in YN14). Note that observational indications for such a strong outflow (see also Schnorr-M\"uller et al.\ 2014) reflects mostly the reduction of accretion rate constrained at large radii near the Bondi radius. Whether such an outflow strength can characterise also the innermost region, where the X-ray radiation is formed, seems to be an open issue. 

We study the dependence on the black hole spin, but mostly we focus on models with a rapidly (but submaximally) rotating black hole, for which the largest differences with respect to previous studies, most of which considered non-rotating black holes, can be expected.

The computational method applied in this work allows to find solutions for a range of bolometric luminosities of $\sim 5 \times 10^{-4}$ to $5 \times 10^{-2}$ of the Eddington luminosity, over which (1) most of the energy provided to electrons is radiated away and (2) the flow structure is not affected by the cooling of protons. This corresponds to the ADAF regime in terms of classification proposed in YN14. In higher luminosity solutions, classified as luminous hot accretion flows (LHAF),  and in lower luminosity solutions, classified as eADAF, condition (2) and (1), respectively, is not satisfied.

The model for black hole binaries, developed by Esin et al.\ (1997, 1998) and a number of later works (see review by Done, Gierli\'nski \& Kubota 2007), relates their spectral changes with changes of both the accretion rate and the truncation distance of a cold disc, the latter being close to the event horizon in the soft state and much farther in the hard spectral state. However, the precise value of the truncation radius in the hard state, which is crucial for the amount of  
the irradiation of an inner hot flow by an outer cold disc, is usually uncertain. In Section \ref{sect:obs} we refer to recent studies indicating that the irradiation is typically weak, except for the initial return from the soft to the hard state, which supports the  comparison of our model with black hole transients. It is not clear whether the above picture can be directly adapted to AGNs, but for those which are considered in Section \ref{sect:obs} we note
substantial observational hints of the lack of an inner cold disc. In our comparison with observations we focus on the spectral index -- Eddington ratio correlation, established  in recent years in large samples of low-luminosity AGNs  (e.g., Gu \& Cao 2009) as well as individual AGNs (e.g., Emmanoulopoulos et al.\ 2012) and black hole binaries (e.g., Wu \& Gu 2008). We also address the value of cut-off energy, $E_{\rm cut}$, however, high-quality measurements are mostly available at luminosities exceeding those of our solutions, at least for a weak direct heating of electrons. 

While our current model involves a comprehensive treatment of both the flow structure and the photon transfer, we make a simplifying assumption that the  power delivered to plasma particles is used to increase their thermal energy. A different approach to modelling the radiation from black-hole systems is used in hybrid, thermal-nonthermal models, which  implement a number of microphysical processes related with the presence of nonthermal particles, in turn, they use crude parametrization of a uniform plasma (see e.g.\ Malzac \& Belmont 2009; Poutanen \& Vurm 2009). The presence of nonthermal electrons can explain a number of observational properties (see e.g.\ Veledina et al.\ 2013 and a recent review in Poutanen \& Veledina 2014). On the other hand, as we discuss in Section \ref{sect:obs}, the comparison of our results with observations puts some constraints on the presence  of such a nonthermal component, although a full investigation of this issue will be possible only in future models implementing the nonthermal processes in the hot flow model.

\section{Hot flow model}
\label{sec:flow}

\begin{figure*} 
\centerline{\includegraphics[height=6.2cm]{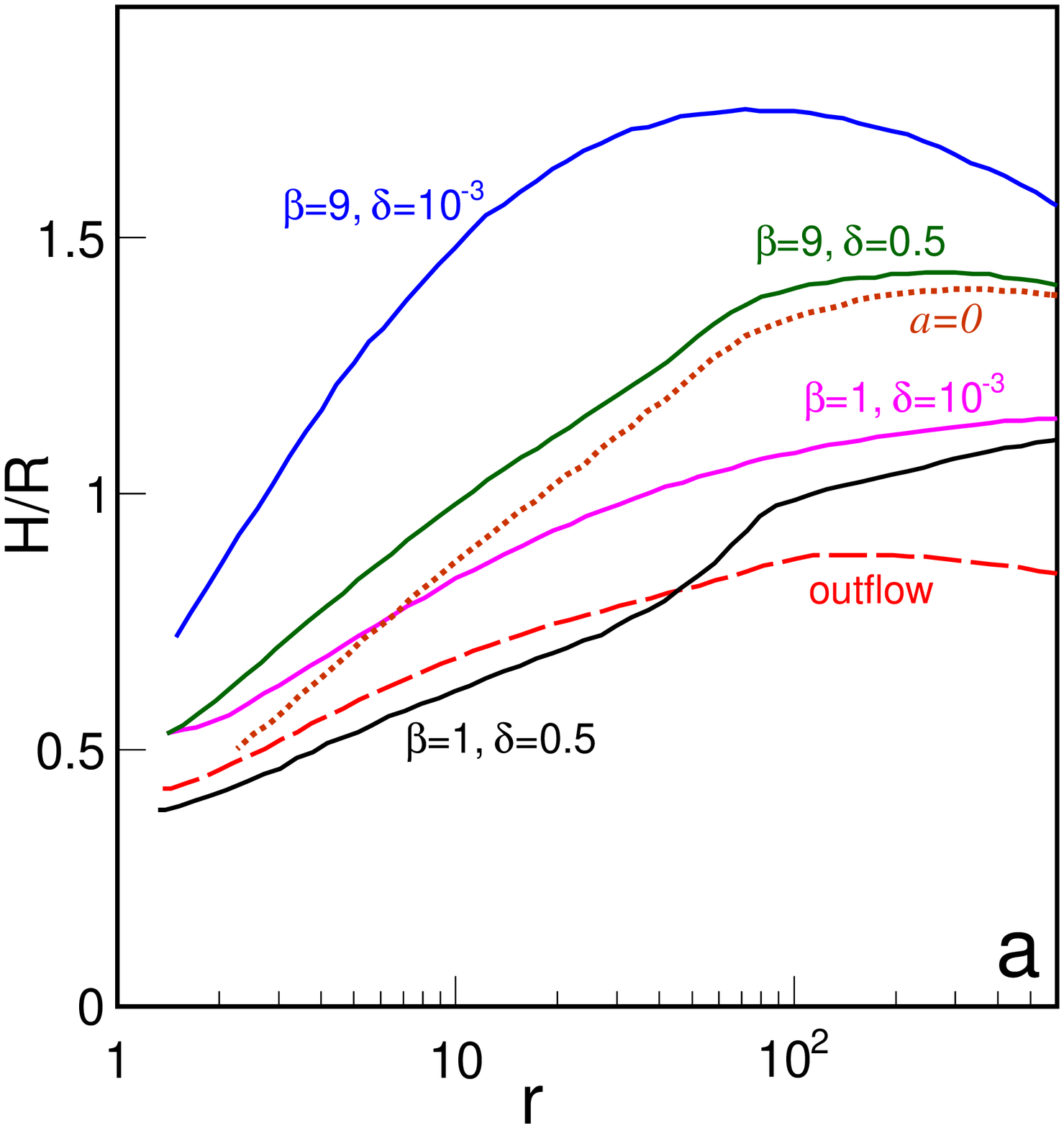}
\includegraphics[height=6.2cm]{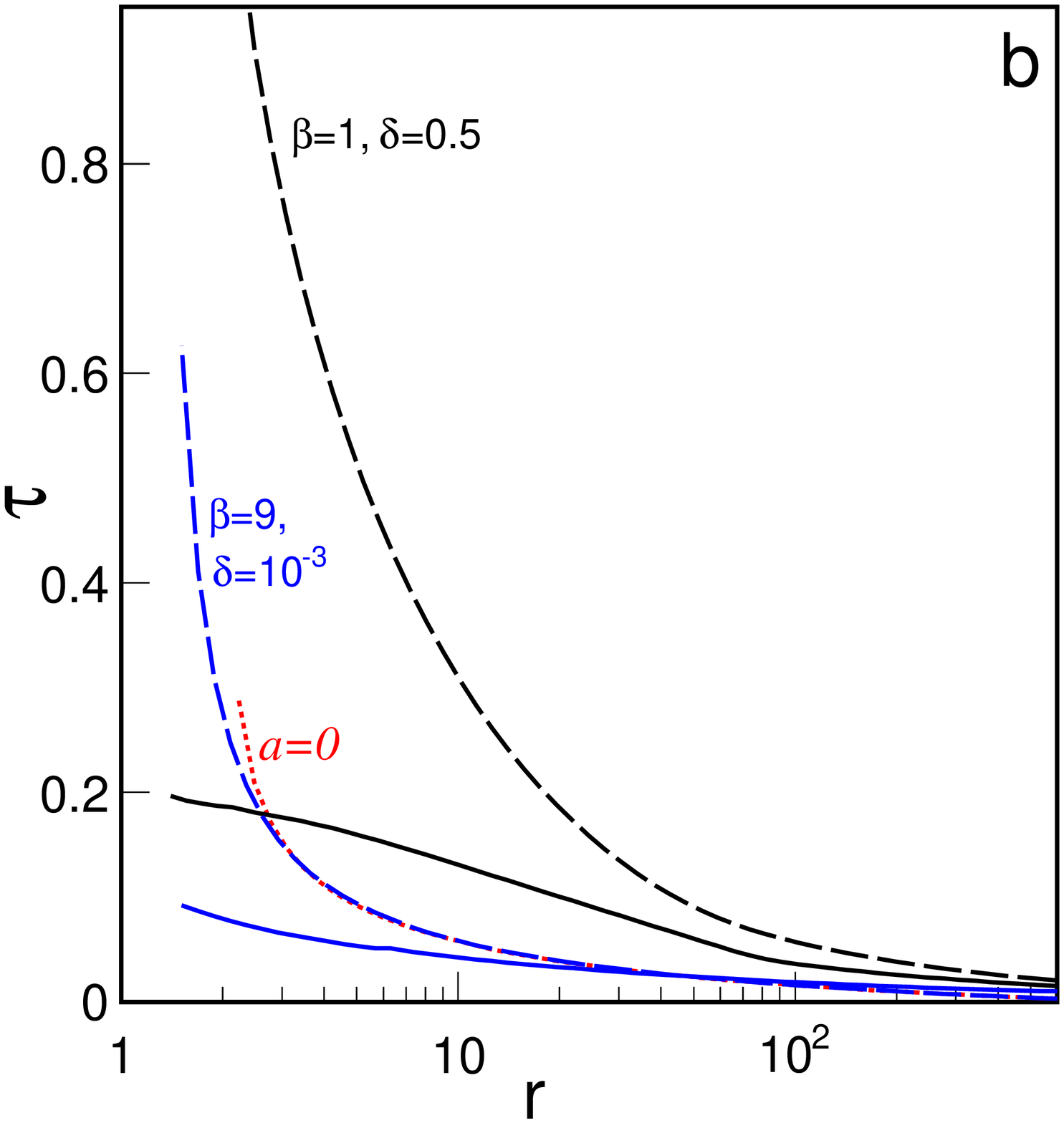} 
\includegraphics[height=6.2cm]{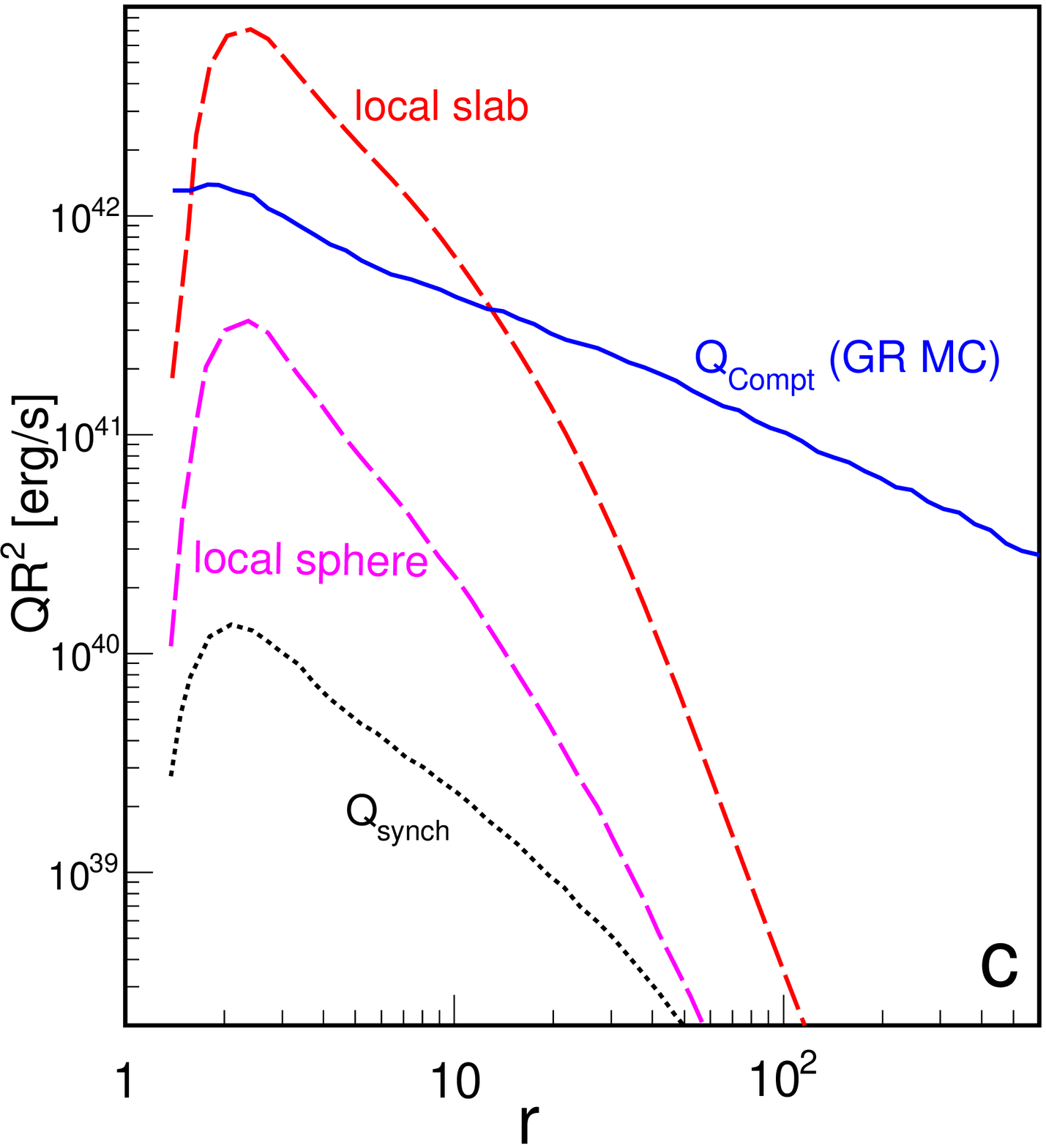}
}
\caption{
Radial profiles of hot flow parameters for $M =2 \times 10^8\, \msun $. All models assume $\dot m=0.1$ and, except for the dotted lines in (a,b), $a=0.95$.
(a) The height-scale profiles; the solid lines show the dependence on  $\beta$ and $\delta$ (as labelled on the figure); the dotted line is for $a=0$ (model a1, see Tables \ref{tab:tab1} and \ref{tab:tab2} for model parameters) and illustrates the (small) dependence on $a$; the dashed line is for the model o2 with an outflow. (b) The Thomson optical depth profiles (see text for definitions). The solid and dashed lines show the vertical and radial (for outgoing photons) depths, respectively. The lower (blue online) lines are for $\beta=9$ and $\delta=10^{-3}$, and the upper (black) are for $\beta=1$ and $\delta=0.5$. The dotted (red) line shows $\tau_r$ for model a1.
(c) The synchrotron (dotted line) and the global Compton (solid line) cooling rates for model a6; the dashed lines show the corresponding local approximation in a sphere and slab geometries. All rates  are vertically integrated, so $QR^2$ gives cooling rates (per unit volume) times volume.
}
\label{fig:flow} 
\end{figure*}

We consider a black hole, characterized by its mass, $M$, and angular momentum, $J$, surrounded by a geometrically thick accretion flow with an accretion rate, $\dot M$. We define the following dimensionless parameters: $r = R / R_{\rm g}$, $a = J / (c R_{\rm g} M)$, $\dot m = \dot M / \medd$, where $\medd= \ledd/c^2$, $R_{\rm g}=GM/c^2$ is the gravitational radius and $\ledd \equiv 4\pi GM m_{\rm p} c/\sigma_{\rm T}$ is the Eddington luminosity. The inclination angle of the line of sight to the symmetry axis is given by $\theta_{\rm obs}$. We assume that  the density distribution is given by $\rho(R,z)=\rho(R,0) \exp(-z^2/2H^2)$, where $H$ is the height scale at $R$ and $z=R \cos \theta$. The ratio of the gas pressure (electron and ion) to the magnetic pressure is denoted by $\beta$; the fraction of the dissipated energy which heats directly electrons is denoted by $\delta$. We also apply the usual assumption that the viscous stress is
proportional to the total pressure, with the proportionality coefficient $\alpha$.

In most cases $\dot m$ does not depend on $r$, but in Section \ref{sect:outfl} we consider  models with an outflow, for which the local accretion rate is given by 
\begin{equation}
\dot m = \dot m_{\rm out} (r/r_{\rm out})^s;
\end{equation}
we assume  $r_{\rm out} = 2 \times 10^4$ and $s=0.3$, for which $\dot m$ at $r=100$ and 2 is reduced by a factor of 5 and $\simeq 16$, respectively, with respect to $\dot m_{\rm out}$.

The model has the following free parameters: $M$, $a$, $\delta$, $\beta$, $\alpha$ and $\dot m$ (unless otherwise specified, the models neglect an outflow, i.e.\ $s=0$). Unless otherwise specified, all models assume $\alpha=0.3$; the dependence on the viscosity parameter is briefly discussed in Section \ref{sect:viscosity}.
For each set of parameters we find the global hydrodynamical solution of the structure  equations. These include the standard conservations of mass, radial momentum and angular momentum, hydrostatic equilibrium and energy equations for electrons and ions. For each equation we use its form given in Manmoto (2000); however, as described in N12, we do not use a  simplifying approximation, ${\rm d}\ln H / {\rm d}\ln R = 1$, adopted in Manmoto (2000).

The solution  yields the radial distribution of the density, $n$, height-scale, $H$, the velocity field, $[v^r,v^{\phi}]$ (we assume that $v^{\theta}=0$), the ion temperature, $T_{\rm i}$, and the initial electron temperature, $T_{\rm e}$. We then assume that the radial distributions of $n$, $v^r$, $v^{\phi}$, $H$ and $T_{\rm i}$ are not affected by changes in $T_{\rm e}$ (see below) and we find the self-consistent electron temperature distribution using the procedure described in N12. We iterate between the solutions of the electron energy equation  and the GR MC Comptonization simulations until we find mutually consistent solutions. 

The electron energy equation is
\begin{equation}
0 = \Lambda_{\rm ie} + \Lambda_{\rm compr} + \delta Q_{\rm vis} - Q_{\rm rad} - Q_{\rm int},
\end{equation}
where $Q_{\rm rad}$ is the radiative cooling rate, $\Lambda_{\rm ie}$ is the electron heating rate by ions via Coulomb collisions, $\Lambda_{\rm compr}$ is the compressive heating rate of electrons, $Q_{\rm vis}$ is the viscous dissipation rate and $Q_{\rm int}$ the rate of advection of the internal energy of
electrons; all rates are defined per unit area of the flow. $Q_{\rm rad}$ includes the self-absorbed synchrotron, $Q_{\rm synch}$, and bremsstrahlung emission rates, and their Comptonization, $Q_{\rm Compt}$; note that for the range of parameters considered here, Comptonization of the synchrotron radiation is always the dominant radiative cooling process. Note also that our $Q_{\rm int}$ term differs from the usual definition in ADAF theory, in which the advection term includes $\Lambda_{\rm compr}$, whereas we consider them separately. Depending on the sign of the temperature gradient, $Q_{\rm int}$ may be negative or positive (representing the release or storing of internal energy) but its absolute value is always smaller, for the considered range of $\dot m$, than both $Q_{\rm rad}$ and at least one of the heating terms ($\Lambda_{\rm ie}$, $\Lambda_{\rm compr}$, $\delta Q_{\rm vis}$) within $r < 100$, so advection of internal energy does not dominate the electron energy balance; this condition determines the lowest $\dot m$ considered here (see below).

We also define  the total heating rate of ions, $Q_{\rm i}^+$,  which includes both the viscous, $(1-\delta) Q_{\rm vis}$, and the compressive heating of ions. Due to high $T_{\rm i}$, the latter, i.e.\ compression work, strongly dominates in models with small $a$. The values of $\Lambda_{\rm ie}$, $\Lambda_{\rm compr}$, $Q_{\rm vis}$, $Q_{\rm i}^+$ and $Q_{\rm rad}$,  integrated over the whole body of the flow, are denoted by $\Lambda_{\rm ie,tot}$, $\Lambda_{\rm compr,tot}$, $Q_{\rm vis,tot}$, $Q_{\rm i,tot}^+$ and $Q_{\rm rad,tot}$, respectively. Due to GR transfer effects, the total power radiated by
the flow, $Q_{\rm rad,tot}$, is higher than the luminosity, $L$, detected far away from the flow.

In this work we present results for $\dot m$ between $\dot m_{\rm down} \simeq 0.01$ and $\dot m_{\rm up} \simeq 0.5$. The specific values of $\dot m_{\rm down}$ and $\dot m_{\rm up}$, for a given set of parameters, result from the following properties. At $\dot m_{\rm up}$, the Coulomb rate, $\Lambda_{\rm ie}$, approaches 10 per cent of $Q_{\rm i}^+$. At larger $\dot m$, our simplified procedure (assuming that the ion energy balance is not significantly affected by the 
exact radial distribution of the electron temperature) is not applicable. At these larger $\dot m$, the full set of the flow structure equations should be solved in each iteration with the MC simulations, which is a complicated procedure (an example of such a solution, but still in the ADAF regime and using a pseudo-relativistic model, is presented in Xie et al.\ 2010). The limiting value of $0.1 Q_{\rm i}^+$ is rough and somewhat uncertain, however, we are not aware of any solutions at relevant $\dot m$ in a GR and global Compton model, from which a more accurate assessment could be inferred.
 
Some models presented below slightly exceed the above constraint. In models a10, s7, s10, s13 and s14 (see Table 1 for model parameters) we get  $0.1 < \Lambda_{\rm ie}/Q_{\rm i}^+ < 0.2$ in some ranges of $r$, so they may be marginally affected by changes of $T_{\rm e}(r)$. In a model with $\delta=10^{-3}$ and $\beta=1$, which predicts relatively small $T_{\rm e}$ and for which we find a remarkable agreement with the observed $\Gamma$--$\lambda_{2-10}$ (Section \ref{sect:obs})
we considered also model s11, with $\dot m=0.6$, in which we get $0.2 < \Lambda_{\rm ie}/Q_{\rm i}^+ < 0.3$ - this solution may slightly deviate from  the fully self-consistent one, in particular, it may slightly overestimate $T_{\rm e}$. In all other models
we get $\Lambda_{\rm ie}/Q_{\rm i}^+ < 0.1$ at all $r$.

Below $\dot m_{\rm down}$, the positive  $Q_{\rm int}$ term starts to dominate over $Q_{\rm rad}$ in most of the radial range and our numerical approach, basing on a dominant role of radiative cooling, is not efficient is finding the self-consistent electron temperature. Particularly for small values of $\delta$, when the energy equation is dominated by two differential terms, $Q_{\rm int}$ and $\Lambda_{\rm compr}$, it is difficult to find a stable solution taking into account the global Comptonization process.

\begin{figure} 
\centerline{
\includegraphics[height=6.34cm]{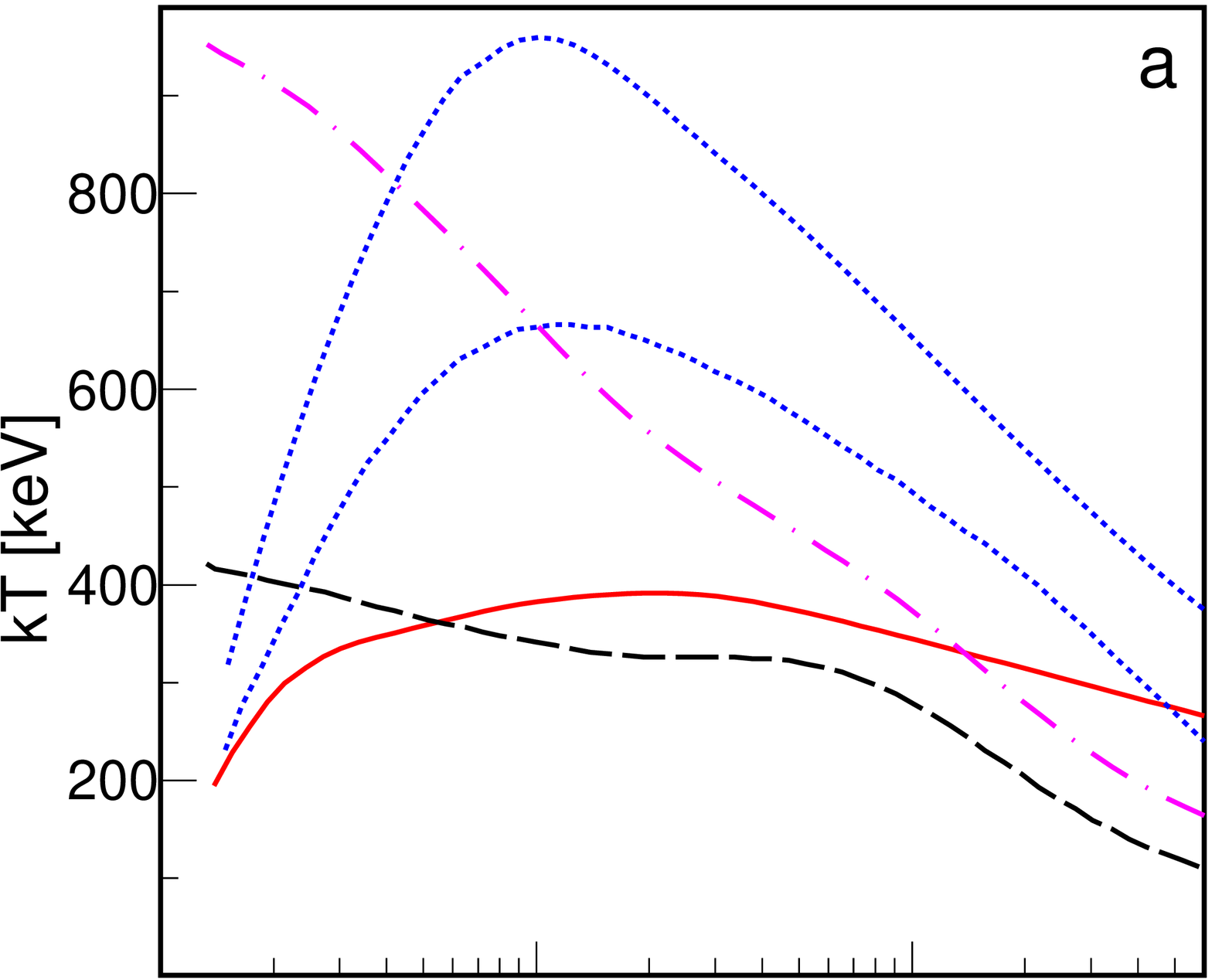}
}
\centerline{
\includegraphics[height=6.6cm]{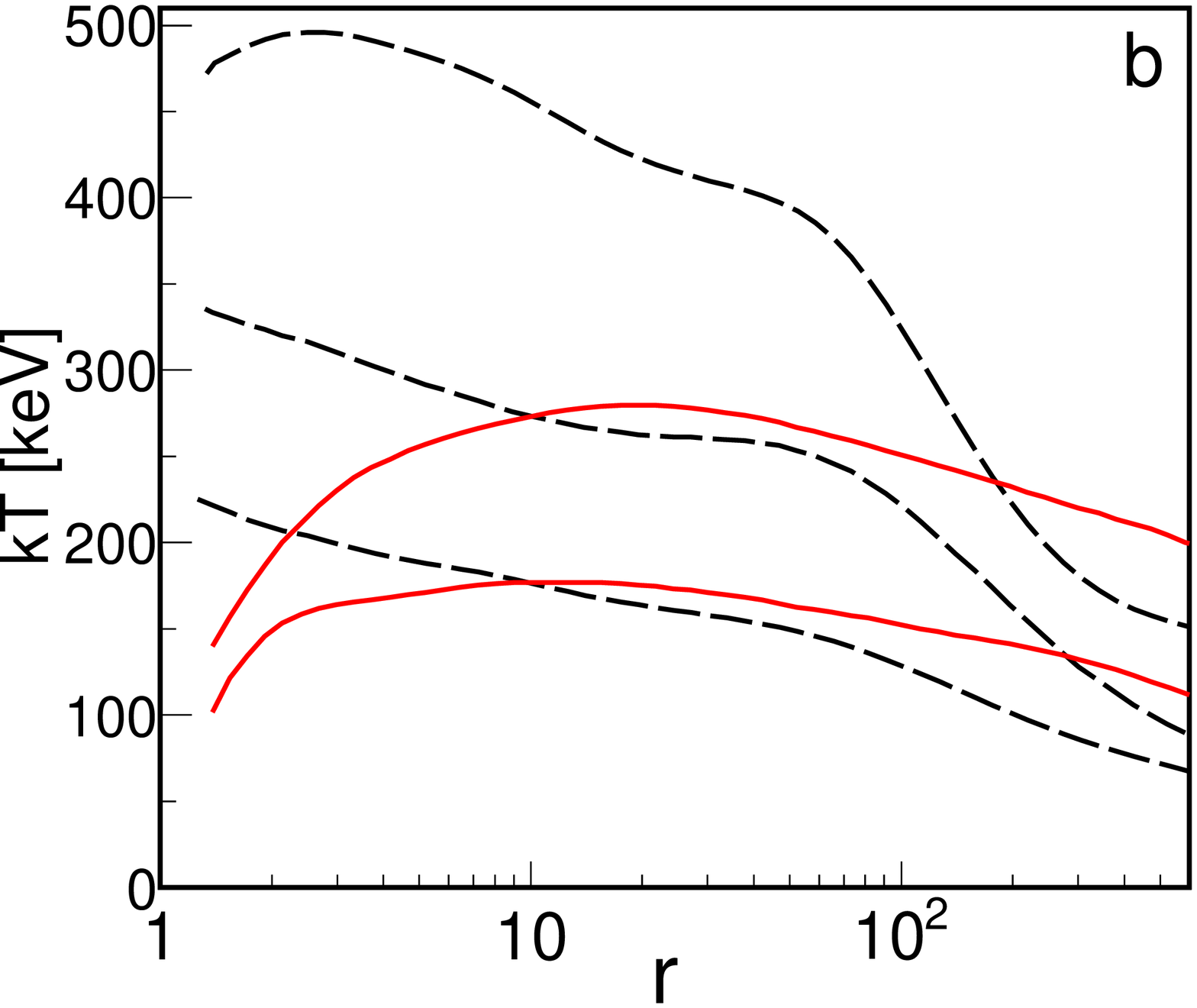}
}
\caption{
Self-consistent electron temperature profiles; all models assume  $a=0.95$. (a) dependence on $M$, $\delta$, $\beta$ and an outflow. The solid (red) and dashed (black) lines are for $\beta=1$ with $\delta=0.5$ (model a9) and $\delta=10^{-3}$  (model a6), respectively. The dotted (blue) lines are for  $\beta=1$ and $\delta=10^{-3}$ with  $M =2 \times 10^8\, \msun $ (model a1; upper) and  $M = 10\, \msun$ (model s1; lower).  The dot-dashed (magenta) line is for  an outflow (model o2). (b) dependence on $\dot m$ for $M = 10\, \msun$; the dashed lines are for $\dot m = 0.03$, 0.1 and 0.3, from top to bottom, in model with $\beta=1$ and $\delta=0.5$; the solid lines are for $\dot m = 0.1$ (upper)  and 0.3 (lower) in model with $\beta=1$ and $\delta=10^{-3}$.  
}
\label{fig:temp} 
\end{figure}

Fig.\ \ref{fig:flow} illustrates properties of hot flow solutions which are important for radiative properties.
In Fig.\ \ref{fig:temp} we show radial distribution of the self-consistent electron temperature. 

Fig.\ \ref{fig:flow}a shows the radial height-scale profiles. Fig.\ \ref{fig:flow}b shows the optical depths of the flow along the vertical and radial directions. The radial optical depth is computed along the radial trajectory, in the equatorial plane, for an outgoing photon, as $\tau_r = \int_r^{r_{\rm out}} \sqrt{\Sigma/\Delta} \gamma (1+|v^r|/c) n_{\rm e} {\rm d}r$, where the first term under integral gives the proper length, in the Kerr metric, along the radial trajectory ($\Sigma = r^2 + a^2$, $\Delta = r^2 - 2r + a^2$) and the second term describes the increase of the probability of scattering in the inflowing plasma (see discussion in Nied\'zwiecki \& Zdziarski 2006); $v^r$ is the radial velocity and $\gamma$ is the Lorentz factor. The vertical optical depth is determined as $\tau_{z} \equiv n_e(r,0) \sigma_{\rm T} H (\pi/2)^{0.5}$. 

An accurate treatment of  the Compton cooling is crucial, as illustrated in Fig.\ \ref{fig:flow}c, where  our global $Q_{\rm Compt}$ is compared with local approximations for the Compton cooling in a sphere and slab geometry given in Dermer et al.\ (1991). The  neglect of the transfer of seed photons from small to large radii strongly underestimates the cooling rate at $r \ga 10$ in any local model. At $r \la 10$, the Compton cooling rate approaches the slab or sphere  approximation depending mainly on $\beta$ (the underlying mechanism is discussed in Section \ref{sect:beta}).
We find that for large $\beta$ (in particular, in all our models with $\beta=9$, except for these with an outflow), $Q_{\rm Compt}$ in the innermost region is reasonably approximated by the local sphere model. For smaller $\beta$ (shown in Fig.\ \ref{fig:flow}c), $Q_{\rm Compt}$ is by a factor of several larger than that of the sphere model. Note, however, that although for $\beta=1$ the spectra of escaping photons are well matched by the slab model with local values of $kT_{\rm e}$ and $\tau_z$ (see Section \ref{sect:compps}), the slab local model overestimates the cooling rate at $2 \la r \la 10$, mostly due to both the presence of the event horizon and only a moderate radial optical thickness ($\tau_r \sim 1$).

\section{Radiative properties}

\label{sect:3}

\begin{figure} 
\centerline{\includegraphics[height=3.4cm]{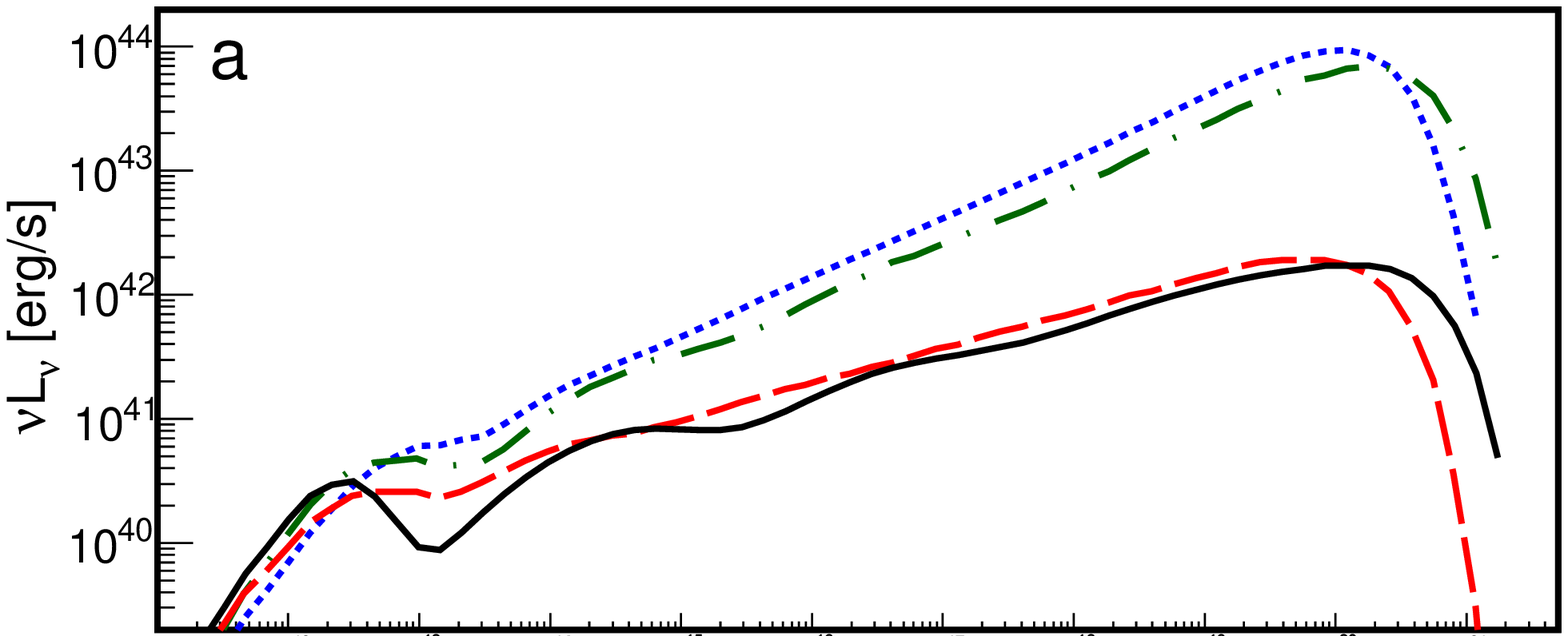}}
\centerline{\includegraphics[height=3.4cm]{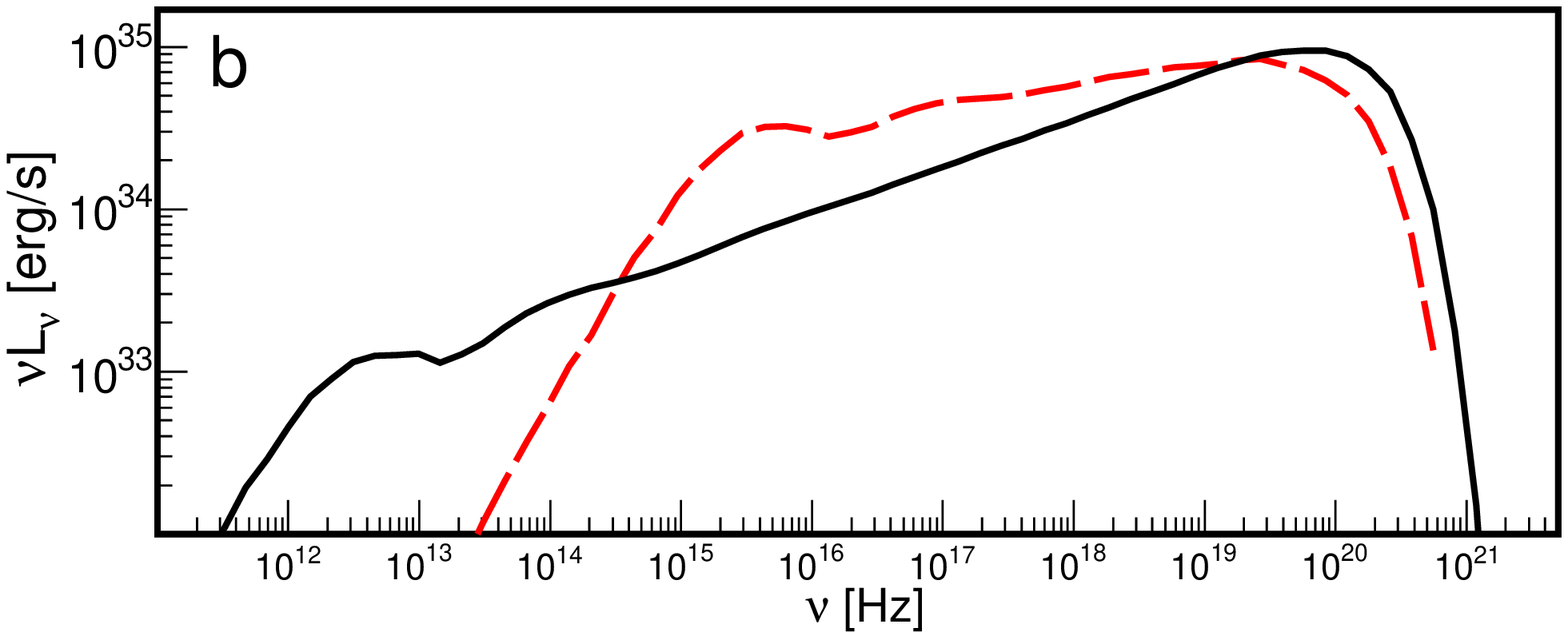}}
\caption{Angle-averaged spectra from MC simulations for $\dot m=0.1$ and $a=0.95$. (a) the dependence on $\delta$ and  $\beta$ in models with  $M =2 \times 10^8 \msun$;  $(\beta,\delta)=(9,10^{-3})$  (solid, black), $(1,10^{-3})$ (dashed, red),  $(9,0.5)$ (dot-dashed, green) and $(1,0.5)$ (dotted, blue). (b) the dependence on $M$ in models with $\beta=1$ and $\delta=10^{-3}$; $M =2 \times 10^8\, \msun$ (solid, black; rescaled by a factor of $0.5\times 10^{-7}$) and $M = 10\, \msun$ (dashed, red).
}
\label{fig:1} 
\end{figure}

Figures \ref{fig:1} and \ref{fig:2}  show our MC spectra of radiation produced in hot flows.  In Tables \ref{tab:tab1} and \ref{tab:tab2} we give the Eddington ratio of the bolometric luminosities  for some characteristic values of the model parameters. In Fig.\ \ref{fig:evol} we show the spectral evolution in the $\Gamma$--$\lambda_{2-10}$ plane,  resulting from the change of $\dot m$, $\delta$ and $\beta$, where $\Gamma$ is the X-ray photon spectral index, $\lambda_{2-10}=L_{2-10}/L_{\rm Edd}$ and $L_{2-10}$ is the luminosity in the 2 to 10 keV range. In some models with low $\dot m$, the emergent spectra have pronounced bumps (as in the bottom spectrum in Fig.\ \ref{fig:2}a), formed by consecutive scattering orders, and the value of the spectral index depends on the energy range. We determine $\Gamma$ in the 2-100 keV range which is more representative for the average slope of the X-ray spectrum; also in our comparison with observations in Section  \ref{sect:obs} we use the observed $\Gamma$ typically measured in such a broad energy range. Some consequences of the departure from a simple power-law are discussed in Section \ref{sect:bump}.  

All results presented in this work correspond to angle-averaged spectra. The flow spectra show a moderate dependence on $\theta_{\rm obs}$, with edge-on ($\cos \theta_{\rm obs} = 0-0.1$) X-ray fluxes typically 2--3 times larger than face-on fluxes ($\cos \theta_{\rm obs} = 0.9-1$). The angular dependence is most relevant for investigation of strength and variability of the reprocessed component and it will be a subject of separate study.

Below we describe simple parametrizations of the model spectra, suitable for comparison with the results of spectral analyses, and then we summarize our conclusions on the dependence on key parameters of the hot flow model.

\subsection{Comparison with COMPPS spectra}
\label{sect:compps}

\begin{figure} 
\centerline{\includegraphics[height=3.72cm]{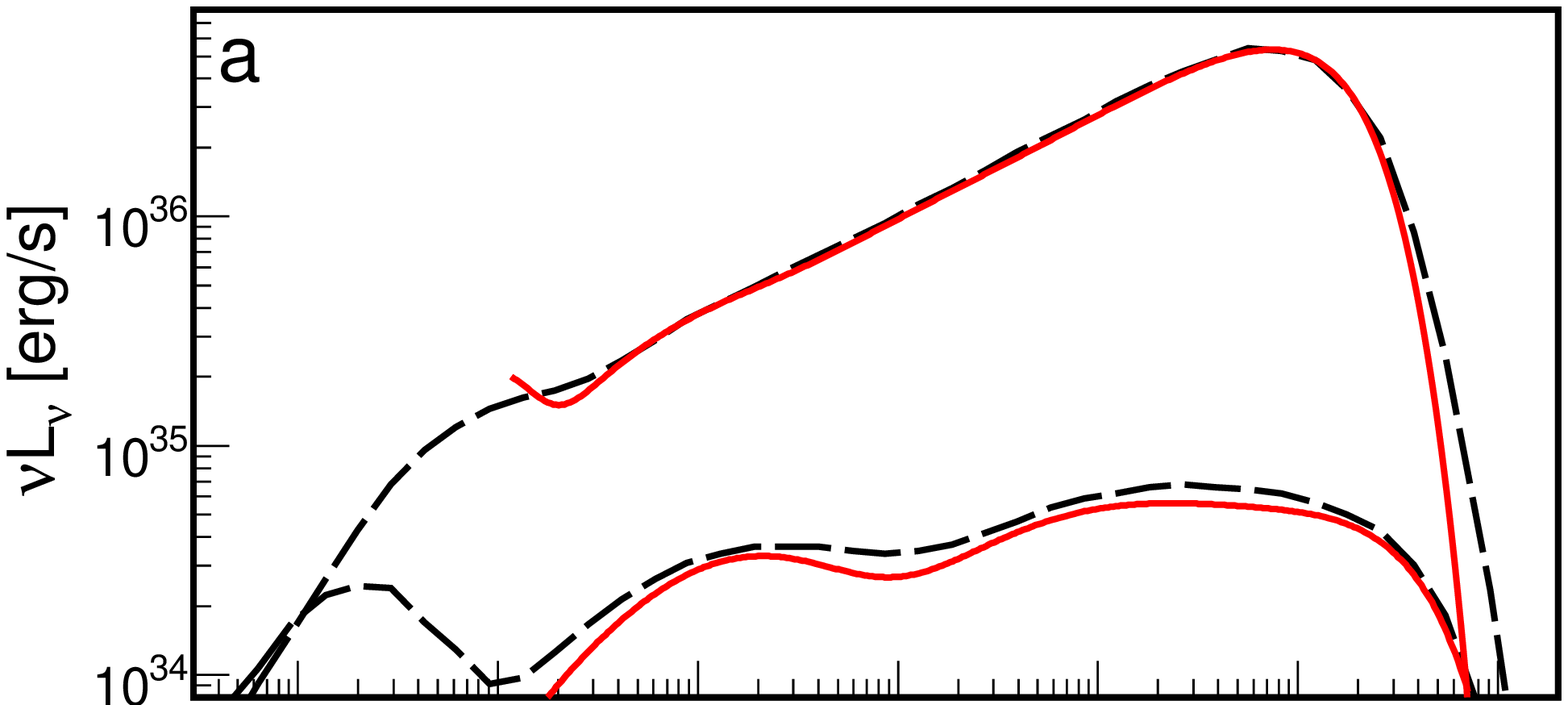}}
\centerline{\includegraphics[height=4.3cm]{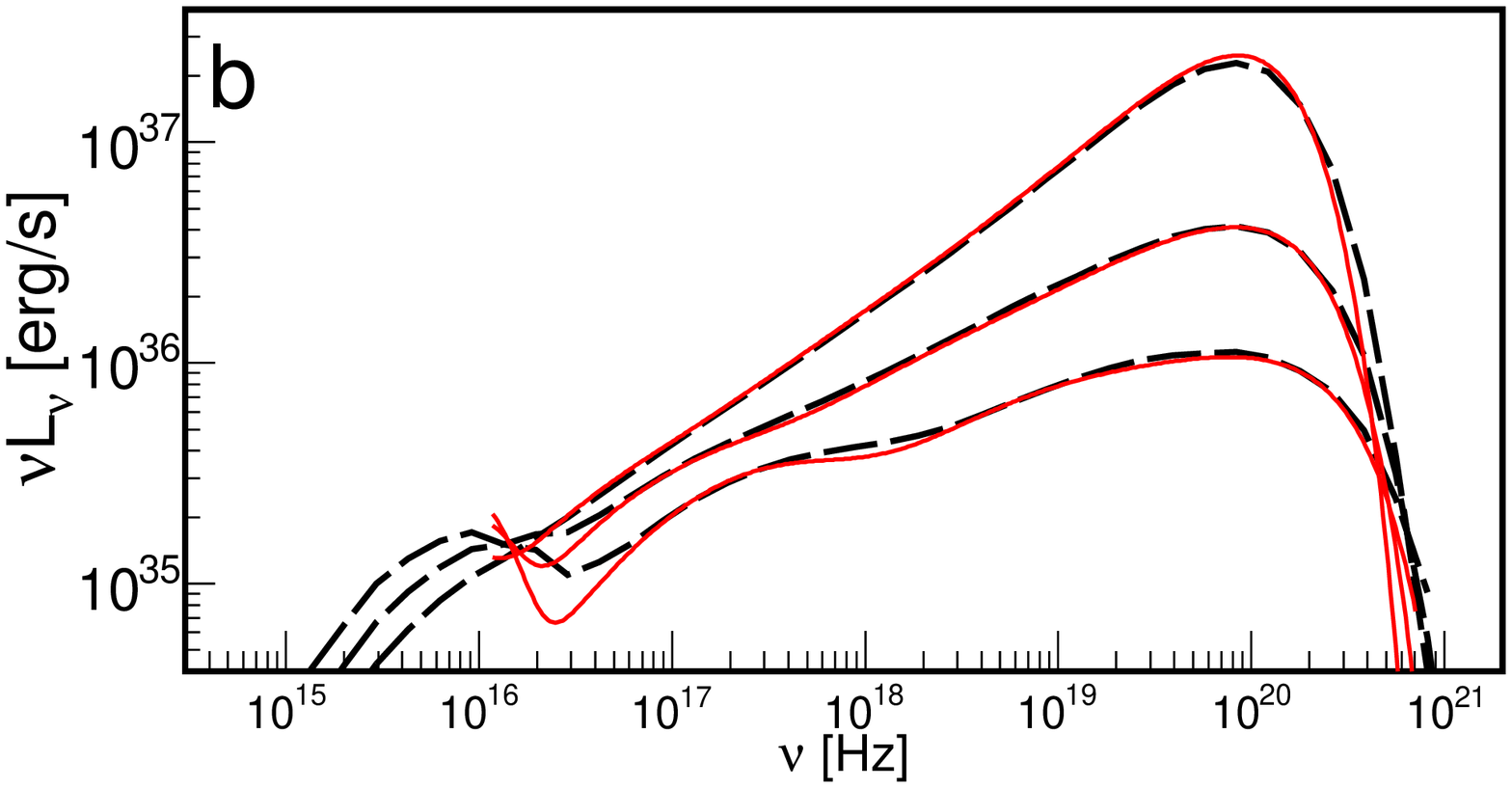}}
\caption{Angle-averaged spectra from MC simulations in models with $a=0.95$ and $M =10\, \msun$ (dashed, black) compared with the best-matching spectra from COMPPS model (solid, red). (a) the dashed spectra are for model s6 (upper) and model s1 (lower); the solid lines are for ($\tau^{\rm PS}, kT_{\rm e}^{\rm PS}$, geometry)=($0.45, 230$ keV, slab) and (0.09, 600 keV, sphere). (b) the dashed spectra are for $\beta=1$, $\delta=0.5$ and $\dot m = 0.033$, 0.1 and 0.3 from bottom to top; the solid lines are for the slab COMPPS model with ($\tau^{\rm PS}$, $kT_{\rm e}^{\rm PS}$) = (0.1, 440 keV), (0.33, 280 keV) and (0.98, 180 keV) from bottom to top.
}
\label{fig:2} 
\end{figure}

The physical models of thermal Comptonization, used in data analysis, typically assume a uniform source, with a simple geometry, characterized by a single electron temperature and a single optical depth. In turn, hot flow solutions are typically characterized by rather complex geometry, with a varying $H/R$ ratio, and broad distributions of both the electron temperature and optical depth. To allow for a comparison with the results of one-zone model fits, we attempted to describe our MC spectra with the high-accuracy Comptonization model, COMPPS (Poutanen \& Svensson 1996); examples are shown in Fig.\  \ref{fig:2}.
The blackbody temperature of  seed photons in COMPPS was fixed to match  the energy of thermal synchrotron peak in the hot flow model.

The plasma parameters of the best-matching COMPPS model, denoted by $\tau^{\rm PS}$ and $T_{\rm e}^{\rm PS}$, typically agree with $T_{\rm e}$ and $\tau$ at the region of the flow, described below by the radial distance, $r_{\rm max}$, which gives the strongest contribution  to the observed spectra. The radial profile of $R^2Q_{\rm Compt}$ always increases toward the horizon, as in Fig.\ 1c. However, GR transfer effects reduce the contribution from the innermost parts and  $r_{\rm max} \simeq 3$--10, depending on the steepness of $Q_{\rm Compt}(r)$. In models with heating of electrons dominated by the compression work,   
the $Q_{\rm Compt}(r)$ profile is flatter and $r_{\rm max} \sim 10$.  A similarly flat  $Q_{\rm Compt}(r)$ occurs for the model with an outflow (for the assumed $s=0.3$) and also here $r_{\rm max} \sim 10$. In flows with large $\delta$ (and $s=0$), as well as those dominated by the Coulomb heating (which occurs at sufficiently large $\dot m$, see Section \ref{sect:mdot} for details), $Q_{\rm Compt}(r)$ is much steeper and $r_{\rm max} \simeq 3$.

It is in general not possible to find the COMPPS spectra exactly matching the flow emission spectra, mostly due to contribution from regions with $T_{\rm e} > T_{\rm e}(r_{\rm max})$. For models with $\tau_r>1$, the COMPPS spectra underpredict the flux at $E > E_{\rm cut}$, and for $\tau_r<1$ the overall spectral shape slightly differs, as illustrated by the top and bottom spectrum, respectively, in Fig.\  \ref{fig:2}a.

Parameters of the best-matching COMPPS model are shown in Fig.\  \ref{fig:comppsparam}.
For some models with $M=2 \times 10^8 \, \msun$ and small $\tau$ (models a1, a8, o3), $T_{\rm e}^{\rm PS} \ga 900$ keV, i.e.\ above the upper boundary of Fig.\ \ref{fig:comppsparam}c.
Spectra of the flows with  the (outward) $\tau_r(r_{\rm max}) \ll 1$ (which occurs only for $\beta=9$ and small $\dot m$) are best-matched by the {\it sphere} COMPPS model (the geometry parameter = -4) with $T_{\rm e}^{\rm PS} \simeq T_{\rm e}(r_{\rm max})$ and $\tau^{\rm PS} \simeq \tau_r(r_{\rm max})$; the slab model gives a much worse approximation of these spectra.

All other models are best-matched by the {\it slab} COMPPS  model (geometry = -1) with $T_{\rm e}^{\rm PS} \simeq T_{\rm e}(r_{\rm max})$ and $\tau^{\rm PS} \simeq 2 \tau_z(r_{\rm max})$. The sphere COMPPS model with the same $T_{\rm e}^{\rm PS}$ typically gives a similar approximation of these spectra, however, its $\tau^{\rm PS}$ parameter is less directly related with parameters of the flow. Specifically, we find $\tau^{\rm PS}{\rm (sphere)} \simeq 2 \tau^{\rm PS}{\rm (slab)}$ for $\tau^{\rm PS}{\rm (slab)} \la 0.5$ and $\tau^{\rm PS}{\rm (sphere)} \simeq 1.5 \tau^{\rm PS}{\rm (slab)}$ for $\tau^{\rm PS}{\rm (slab)} \ga 1$.

\subsection{Magnetic field strength}
\label{sect:beta}

As seen in Fig.\ \ref{fig:1}a, the major spectral change resulting from the increase of $\beta$ concerns the cut-off energy. Flows with smaller $\beta$ have much smaller $T_{\rm e}$, e.g.\ for $M=2 \times 10^8 \, \msun$, $\beta=1$ gives $kT_{\rm e}$ smaller by 100--500 keV than $\beta=9$, see Fig.\ \ref{fig:temp}. A similar in magnitude difference occurs between $T_{\rm e}^{\rm PS}$ for the observed spectra, see Fig.\  \ref{fig:comppsparam}a.  This is in a qualitative agreement with the conclusion of Esin et al.\ (1998), who attributed the decrease of $T_{\rm e}$ with decreasing $\beta$ to increased synchrotron emissivity.
We note that the obvious dependence of $Q_{\rm synch}$ on $\beta$ is insufficient for significant changes of $T_{\rm e}$, because of the extreme sensitivity of both $Q_{\rm synch}$ and $Q_{\rm Compt}$ on $T_{\rm e}$. Below we discuss in some details the actual  mechanism leading  to a  reduction of $T_{\rm e}$ in small-$\beta$ flows.

As we see in Fig.\ 1a, flows with a smaller $\beta$ are geometrically thinner, see also  figure 4 in Popham \& Gammie (1998). The underlying physical mechanisms are  discussed by Popham \& Gammie (1998) and Quataert \& Narayan (1999). Namely, in flows with smaller $\beta$ (larger $B$) a larger fraction of the accretion power is used to build up the magnetic field strength, therefore, the energy heating the particles, and hence the ion temperature and pressure are smaller. The contribution of magnetic field to the total pressure is characterized by the  adiabatic index ($4/3$) smaller than that of the ion gas ($\simeq 5/3$; ions are non-relativistic through most of the flow). Then, a flow with larger $\beta$ is less compressible and has a larger height-scale.

From the continuity equation, $n \propto  (H |v^r|)^{-1}$, and because $|v^r| \propto T_{\rm i}  H$ (cf.\ Frank et al.\ 2002), the product of $n H$ increases when $H$ decreases; therefore, $\tau_{z}$ is larger in models with smaller $\beta$, see Fig.\ 1b. Furthermore, the increased density results in an even larger increase of the radial optical depth in these models. This, in turn, implies that the  $\tau_r/\tau_z$ ratio is larger in flows with smaller $\beta$, which makes such flows closer to a slab. 

For $\beta=1$, $T_{\rm e}$ is typically smaller by a factor of $\sim 2$, while $B$ larger by a factor of $\sim 4$,  than for $\beta=9$. 
Such combinations of $B$ and $T_{\rm e}$ yield similar  $Q_{\rm synch}$, i.e.\ the seed photon fluxes are equal for both values of $\beta$. If the flow structure remained unchanged, $Q_{\rm Compt}$ would be by an order of magnitude smaller for $\beta=1$ (due to twice smaller $T_{\rm e}$). We then reiterate  that the change of geometry  leading to  a much larger $\tau_r$ (making the flow closer to a slab), is the key mechanism compensating the decrease of $T_{\rm e}$ in models with smaller $\beta$.

\begin{table}
\begin{center}
\begin{tabular}{|llllll|l|}
\hline
\multicolumn{1}{|c|}{model}  & 
 \multicolumn{1}{|c|}{$M$}  & 
 \multicolumn{1}{|c|}{$\dot m$ } & 
\multicolumn{1}{|c|}{$a$ } & 
 \multicolumn{1}{|c|}{$\delta$ } & 
 \multicolumn{1}{|c|}{$\beta$ } &  
 \multicolumn{1}{|c|}{$L/L_{\rm Edd}$ }   
\\
&
 \multicolumn{1}{|c|}{[$\msun$]}  & 
  & 
  & 
  & 
  &  
 \multicolumn{1}{|c|}{[\%]} 
\\
\hline
\hline
a1 & $2 \times 10^8$ & 0.1 &  0 & $10^{-3}$ & 9 & 0.04 
\\
\hline
a2 & $2 \times 10^8$ & 0.1 &  0 & 0.5   & 9 &  0.24  
 \\
\hline
a3 & $2 \times 10^8$ & 0.1 & 0.95 & $10^{-3}$ & 9 & 0.04 
\\
\hline
a4 & $2 \times 10^8$ & 0.3 & 0.95 & $10^{-3}$ & 9 & 0.12 
\\
\hline
a5 & $2 \times 10^8$ & 0.1 & 0.95 & 0.5   & 9 & 0.8 
 \\
\hline
a6 & $2 \times 10^8$ & 0.1 & 0.95 & $10^{-3}$ & 1 & 0.04 
  \\
\hline
a7 & $2 \times 10^8$ & 0.3 & 0.95 & $10^{-3}$ & 1 & 0.28 
\\
\hline
a8 & $2 \times 10^8$ & 0.01 & 0.95 & 0.5 & 1 & 0.14
  \\
\hline
a9 & $2 \times 10^8$ & 0.1 & 0.95 & 0.5 & 1 & 1.4 
  \\
\hline
a10 & $2 \times 10^8$ & 0.3 & 0.95 & 0.5 & 1 & 4.7 
\\
\hline
a11 & $2 \times 10^8$ & 0.1  & 0.998 & $10^{-3}$ & 9 & 0.04 
 \\
\hline
a12 & $2 \times 10^8$ & 0.1  & 0.998 & 0.1 & 9 & 0.12 
 \\
\hline
a13 & $2 \times 10^8$ & 0.1 &  0.998 & 0.5 & 9 & 1 
\\
\hline
s1 & 10 & 0.1  & 0.95 & $10^{-3}$ & 9 & 0.04 
\\
\hline
s2 & 10 & 0.3  & 0.95 & $10^{-3}$ & 9 & 0.09
\\
\hline
s3 & 10 & 0.6  & 0.95 & $10^{-3}$ & 9 & 0.28
\\
\hline
s4 & 10 & 0.1  & 0.95 & 0.5 & 9 & 0.7 
\\
\hline
s5 & 10 & 0.1  & 0.95 & 0.5 & 1 & 1.4 
\\
\hline
s6 & 10 & 0.1  & 0.95 & 0.5 & 0.43 & 1.6 
\\
\hline
s7 & 10 & 0.3  & 0.95 & 0.5 & 1 & 4.8 
\\
\hline
s8 & 10 & 0.1  & 0.95 & $10^{-3}$ & 1 & 0.05 
\\
\hline
s9 & 10 & 0.3  & 0.95 & $10^{-3}$ & 1 & 0.27 
\\
\hline
s10 & 10 & 0.45  & 0.95 & $10^{-3}$ & 1 & 0.72 
\\
\hline
s11 & 10 & 0.6  & 0.95 & $10^{-3}$ & 1 & 1.5 
\\
\hline
s12 & 10 & 0.1  & 0.95 & $10^{-3}$ & 0.3 & 0.04 
\\
\hline
s13 & 10 & 0.3  & 0.95 & $10^{-3}$ & 0.3 & 0.35 
\\
\hline
s14 & 10 & 0.5  & 0.95 & $10^{-3}$ & 0.3 & 1
\\
\hline
v1 & 10 & 0.1 &  0.95 & $10^{-3}$ & 1 & 0.16 
\\
\hline
\end{tabular}
\end{center}
\caption{\small 
Model parameters and the Eddington ratio of bolometric luminosity for models with constant $\dot m$. All models assume $\alpha = 0.3$, except for model v1 which is for $\alpha = 0.1$.
}    
\label{tab:tab1}
\end{table}

In Sections \ref{sect:delta}, \ref{sect:outfl} and \ref{sect:turn} we refer to the above discussion, as the increase of $Q_{\rm Compt}$, in flows approaching the slab geometry, occurs also due to the presence of outflow (reducing the compressive heating of ions) as well as the direct heating of electrons (reducing the viscous heating of ions), and we may expect it also at large $\dot m$ (due to increased Coulomb cooling of ions).

Whereas the decrease of $\beta$ from 9 to 1 results in a strong reduction of $T_{\rm e}^{\rm PS}$, a further decrease to super-equipartition values, $\beta < 1$, has a moderate effect, see Fig.\ \ref{fig:comppsparam}a. The $\Gamma$--$\lambda_{2-10}$ relation is weakly affected by the change of $\beta$, see Fig.\ \ref{fig:evol}. We also note that the height-integrated pressure is larger in models with smaller $\beta$ which then results in higher dissipation rate, $Q_{\rm vis} \propto H p_{\rm tot}$, in these models. 

\subsection{Efficiency of the viscous heating of electrons}
\label{sect:delta}

\begin{figure} 
\centerline{\includegraphics[width=8cm]{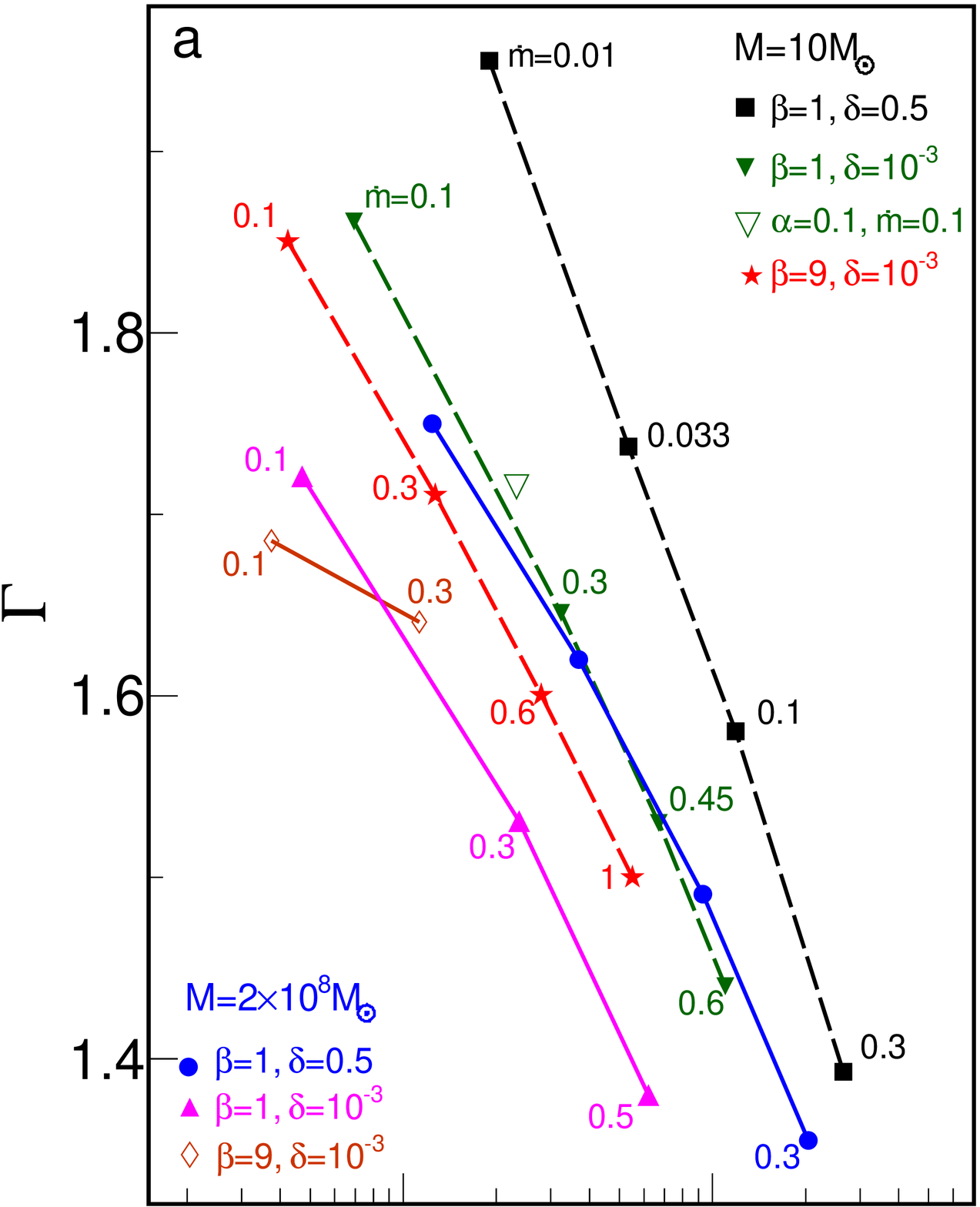}}
\centerline{\includegraphics[width=7.99cm]{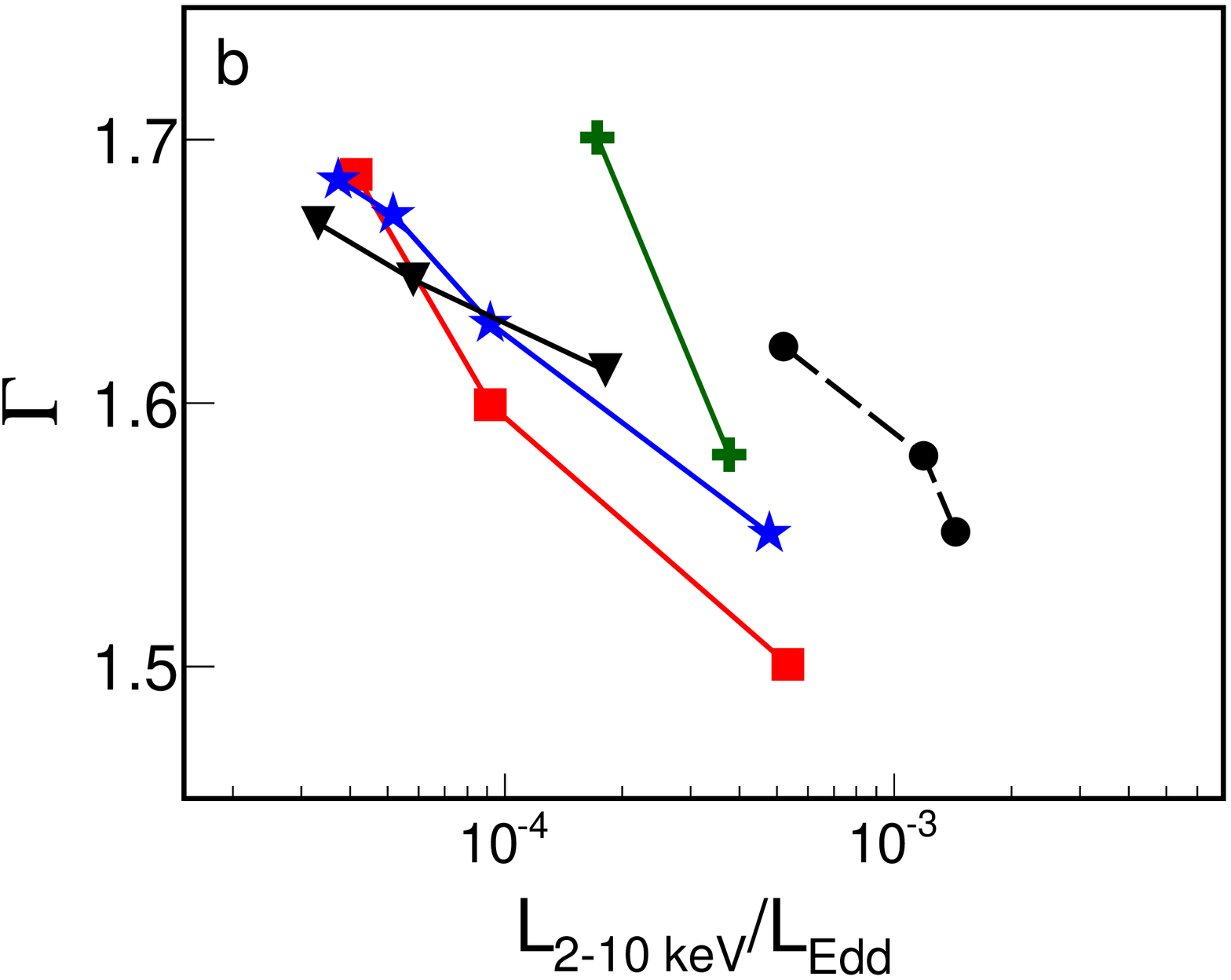}}
\caption{Spectral evolution, in the $\Gamma$--$\lambda_{2-10}$ plane, resulting from the change of $\dot m$, $\delta$ and $\beta$; the model points connected by solid and dashed lines are for $M =2 \times 10^8\, \msun $ and $M = 10\, \msun $, respectively. (a) each set of points corresponds to a fixed $\beta$ and $\delta$, and $\dot m$ varies as indicated in the figure (for circles with solid line: $\dot m = 0.01$, 0.033, 0.1, 0.3 from left to right); all models assume $a=0.95$. The open triangle is for $\alpha=0.1$ (model v1).
(b) the (black) circles are for $\beta = 0.43$, 1 and 9 (models s4, s5, s6) from left to right; (green) crosses are for an outflow with $\dot m_{\rm out} = 0.17$ (model o3, upper) and 0.5 (model o2, lower). The remaining points illustrate the change of $\delta$ for $\beta = 9$, $\dot m = 0.1$ and various $a$:
(black) triangles  are for $a=0$ with $\delta=10^{-3}$, 0.1 and 0.5; (blue) stars are for $a=  0.95$ with $\delta=10^{-3}$, 0.01, 0.1 and 0.5; (red) squares are for $a = 0.998$ with $\delta=10^{-3}$, 0.1 and 0.5. 
}
\label{fig:evol} 
\end{figure}

Fig.\ \ref{fig:1}a illustrates the increase of luminosity and spectral hardening resulting from the increase of $\delta$. Fig.\ \ref{fig:evol}b shows the changing $\delta$ effect in the $\Gamma$--$\lambda_{2-10}$ plane. Regardless of the value of $a$, in models with $\delta=10^{-3}$ the direct heating of electrons has a minor effect, with at most 10 per cent contribution to the total heating rate, whereas in models with $\delta=0.5$ the direct heating dominates the total heating of electrons. Then, the increase of $\delta$ from $10^{-3}$ to 0.5 results in the increase of $L$ by a factor of several for small $a$ and by over an order of magnitude for large $a$. For $a=0.998$, $\delta = 0.1$ gives $L$ by  a factor of $\simeq 3$  larger than $\delta=10^{-3}$. 

The increase of the $\delta$ parameter implies the decrease of the ion heating and, because the energy heating electrons is mostly radiated away and hence electrons contribute negligibly to the total pressure, it results in the decrease of $H/R$.  However, the related enhancement of the Compton cooling is outweighed by the increase of heating and the flow temperature at $r<10$ increases with increasing $\delta$, see Fig.\ \ref{fig:temp}b. 

The dependence of  $L$ on $\delta$ indicates that for $\delta=0.5$ the same $L$ is produced at roughly an order of magnitude smaller $\dot m$ than for $\delta = 10^{-3}$; this implies an order of magnitude smaller $\tau^{\rm PS}$ for large $\delta$, see Fig.\ \ref{fig:taulum}. Hence, $kT_{\rm e}^{\rm PS}$ is much higher (by $\sim 400$ keV) in models with $\delta = 0.5$ at a given $L$, see Fig.\ \ref{fig:comppsparam}a. Also as a result of larger $T_{\rm e}$ (implying a much stronger seed photon flux), flows with $\delta= 0.5$ produce softer spectra, with $\Delta \Gamma \simeq 0.2$, than flows with $\delta = 10^{-3}$ at the same $\lambda_{2-10}$, see Fig.\ \ref{fig:evol}a.

We note hints that a transition to a cold disc at large $\dot m$ may proceed differently in flows with large and small $\delta$. Due to the decrease of the ion heating rate, for $\delta=0.5$ the  $\Lambda_{\rm ie}/Q_{\rm i}^+$ ratio is larger by a factor of $\sim 2$--3 than in models with $\delta=10^{-3}$
(at the same $\dot m$). Furthermore, for $\delta=0.5$ the largest $\Lambda_{\rm ie}/Q_{\rm i}^+$ ratio occurs at $r \sim 20$, implying that  the transition should occur in the inner flow first. In contrary, the monotonic increase of $\Lambda_{\rm ie}/Q_{\rm i}^+$  with increasing  $r$ for $\delta=10^{-3}$ indicates that in small-$\delta$ flows the transition occurs only beyond some transition radius, $r_{\rm tr}$, and the value of $r_{\rm tr}$ progressively decreases with increasing $\dot m$.

\begin{table}
\begin{center}
\begin{tabular}{|llllll|l|}
\hline
\multicolumn{1}{|c|}{model}  & 
 \multicolumn{1}{|c|}{$M$}  & 
 \multicolumn{1}{|c|}{$\dot m_{\rm out}$ } & 
\multicolumn{1}{|c|}{$a$ } & 
 \multicolumn{1}{|c|}{$\delta$ } & 
 \multicolumn{1}{|c|}{$\beta$ } &  
 \multicolumn{1}{|c|}{$L/L_{\rm Edd}$ }   
\\
&
 \multicolumn{1}{|c|}{[$\msun$]}  & 
  & 
  & 
  & 
  &  
 \multicolumn{1}{|c|}{[\%]} 
\\
\hline
\hline
o1 & $2 \times 10^8$ & 0.5 &  0.95 & 0.5 & 1 & 0.7 
\\
\hline
o2 & $2 \times 10^8$ & 0.5 &  0.95 & 0.5 & 9 & 0.5 
\\
\hline
o3 & $2 \times 10^8$ & 0.17 & 0.95 & 0.5 & 9 & 0.29 
\\
\hline
o4 & $10$ & 0.5 & 0.95 & 0.5 & 9 & 0.5 
\\
\hline
o5 & $10$ & 0.5 & 0.95 & 0.5 & 1 & 0.7 
\\

\hline
\end{tabular}
\end{center}
\caption{\small 
Same as Table \ref{tab:tab2} but for models with $\dot m = \dot m_{\rm out}(r/r_{\rm out})^{0.3}$.
}    
\label{tab:tab2}
\end{table}

\subsection{Black hole spin}
\label{sect:spin}

The spin affects the X-ray emission mostly through the impact on the heating rate. The increase of $a$ stabilizes the circular motion of the flow and hence it results in the increase of the viscous heating rate. For large $\delta$, this leads to a strong dependence of the flow luminosity on $a$; e.g.\ for $\delta=0.5$,  $L$ is by a factor of 5 larger  at $a=0.998$ than at $a=0$. The submaximal $a=0.95$ gives $L$ smaller by only $\sim 20$ per cent than  $a=0.998$; the difference of $Q_{\rm vis,tot}$ between these two $a$ is much larger, with a factor of $\sim 2$--3, however, the difference between $L$ is reduced by GR transfer effects (most of the dissipation for $a=0.998$ occurs close to the event horizon).

We also note a factor of $\sim 4$--5 stronger Coulomb heating, resulting from a larger density and ion temperature at $r<10$, for large $a$. 
On the other hand, the impact of $a$ on the compressive heating is insignificant. Therefore, for small $\delta$ the difference between $L$ for different $a$ increases with increasing $\dot m$ (i.e.\ increasing relative contribution of Coulomb heating).

As we see in Fig.\ 1a, $a$ has a much weaker influence on $H/R$ than $\delta$ or $\beta$. The density, and thus $\tau_z$, is slightly larger in models with large $a$ within the innermost several $R_{\rm g}$ (by a factor of $\sim 1.5$ at $r=3$) as a result of $|v^r|$ being smaller. However, the increased kinematic term for outgoing electrons compensates the decrease of density for smaller $a$, which reduces the dependence of $\tau_r$ on $a$, see Fig.\ 1b.

Finally, the total heating rate of ions is significantly  larger in models with large $a$, however, the difference between large and small $a$ occurs only within $r<10$. At $r>10$, the $\Lambda_{\rm ie}/Q_{\rm i}^+$ ratio is independent of $a$. Therefore, the value of $r_{\rm tr}$ at a given $\dot m$ should not depend on $a$,  until it decreases down to $r_{\rm tr} \sim 10$. This, with the dependence of $L$ on $a$, implies that the luminosity of the flow with a given $r_{\rm tr}$ should be $\sim 5$ times larger for high $a$. We may also expect that flows with large $a$ can reach much larger maximum luminosities due to the strong dependence of $Q_{\rm vis}$ on $a$.

\subsection{Accretion rate; radiative efficiency}
\label{sect:mdot}

Fig.\ \ref{fig:2}b shows examples of the change of spectra  due to the change of $\dot m$. The amount of the increase of $L$ with $\dot m$  depends on the  process which dominates the electron heating, as discussed below; we also describe here the related scaling of the radiative efficiency, defined as $\eta \equiv  L/(\dot M c^2)$, with $\dot m$.

\begin{figure*} 
\centerline{\includegraphics[height=8.8cm]{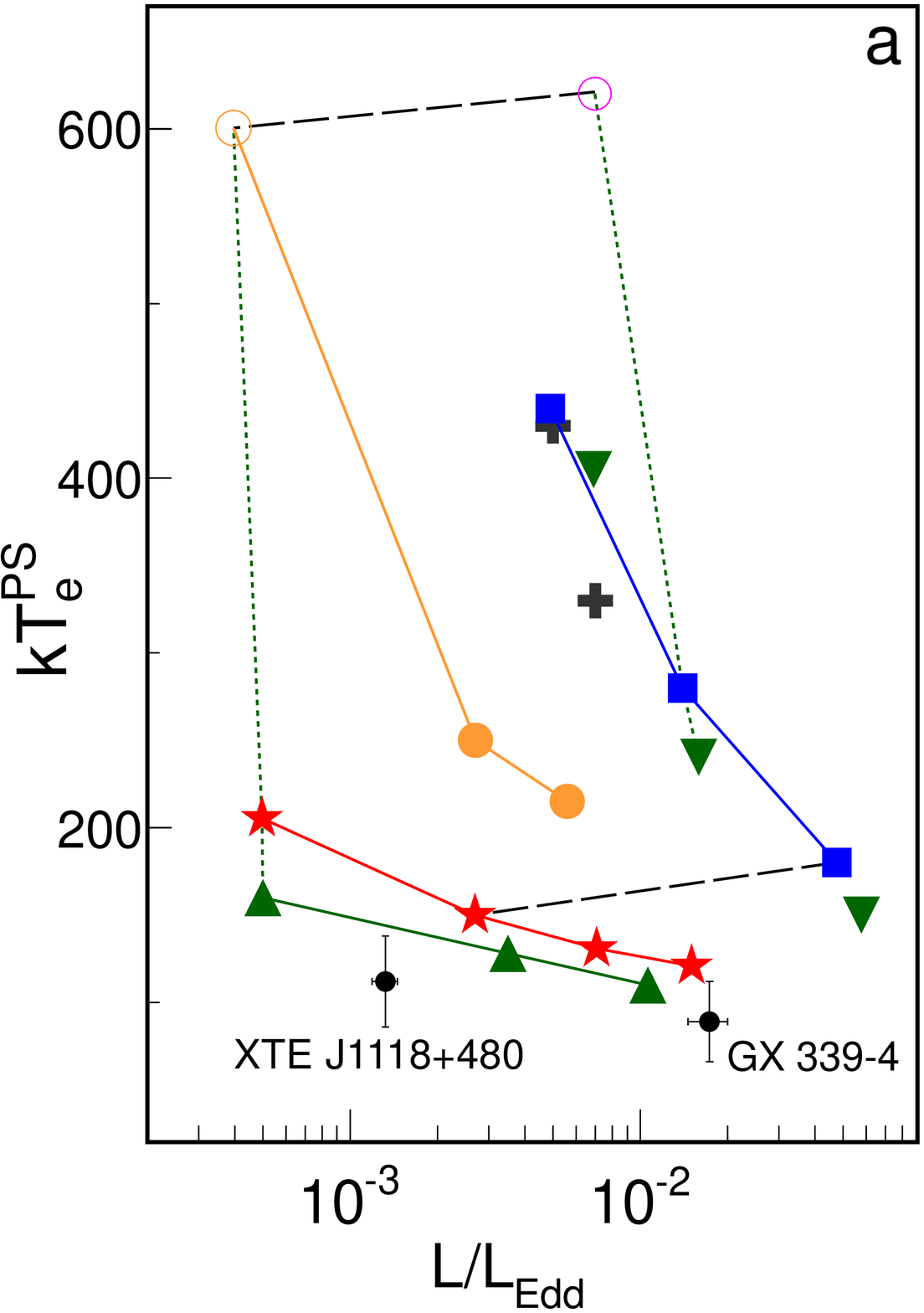}\includegraphics[height=8.8cm]{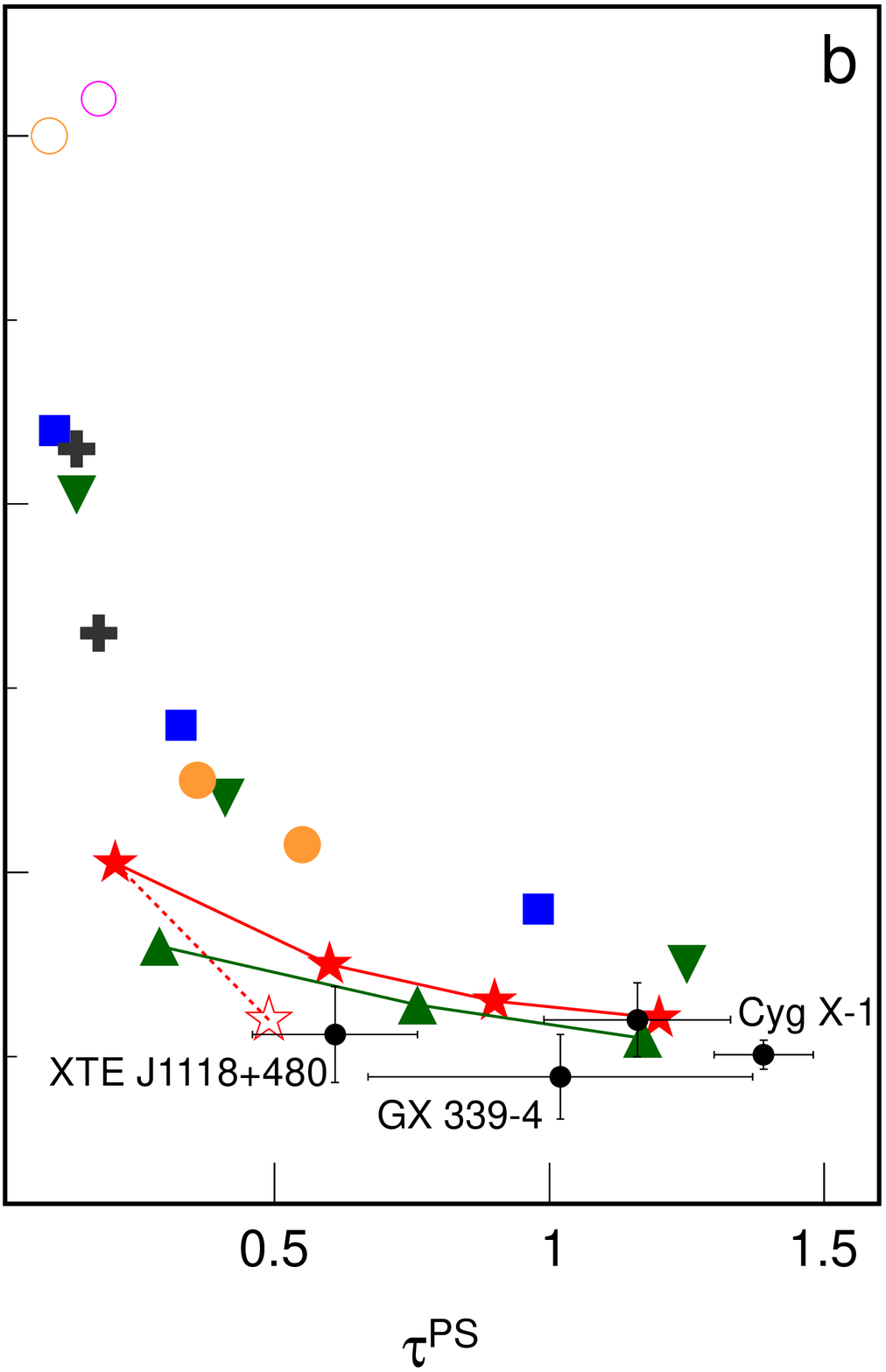}\includegraphics[height=8.8cm]{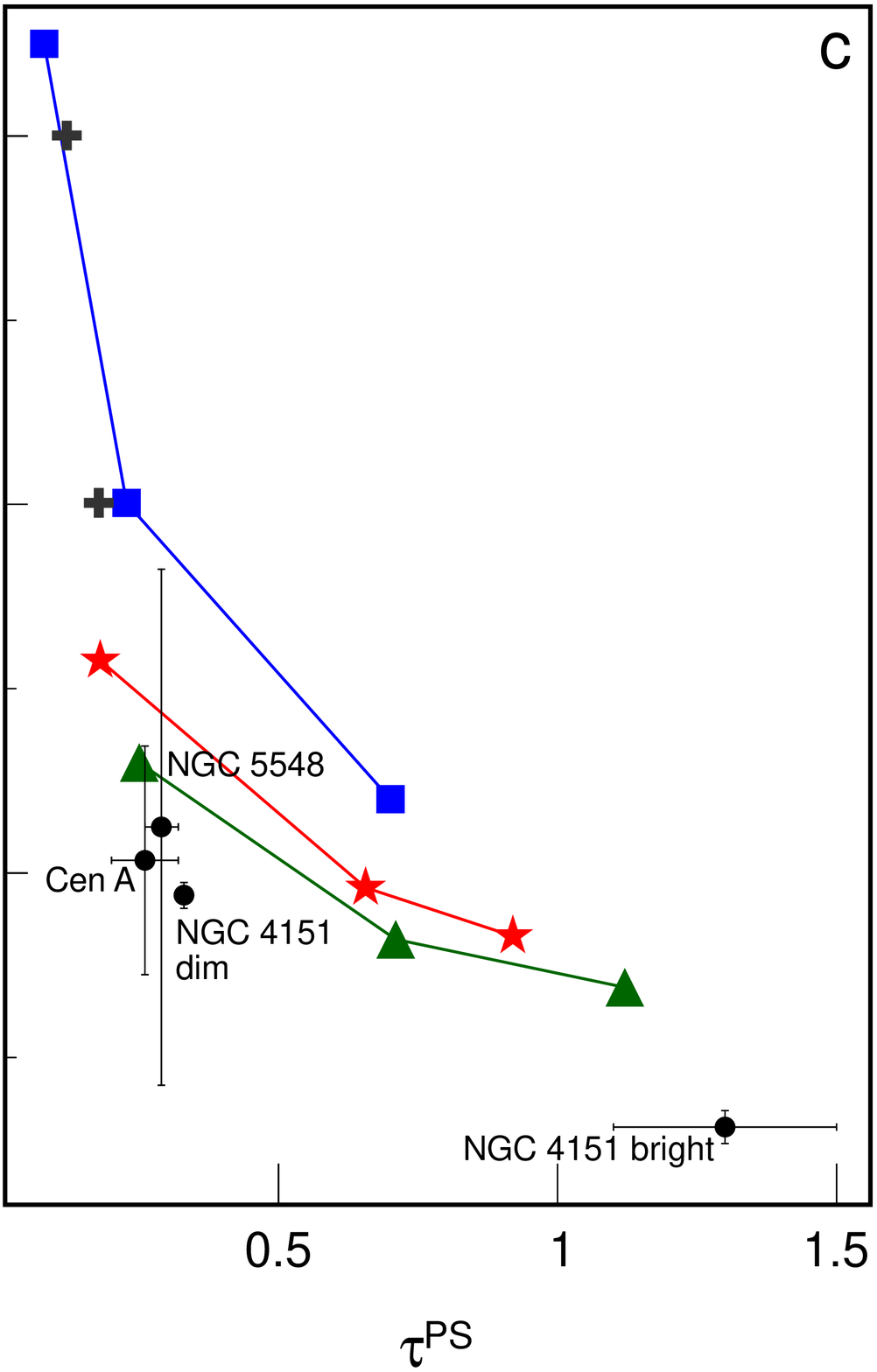}}
\caption{In all panels, the model points show parameters, $T_{\rm e}^{\rm PS}$ and  $\tau^{\rm PS}$, of the COMPPS model  best-matching our MC spectra of a hot flow; all correspond to the slab model, except for the two open  circles (for $\beta=9$ and $\dot m=0.1$) in the top  of panels (a) and (b), which are for the sphere. Panels (a) and (b) are for $M =10\, \msun$ and panel (c) is for $M =2 \times 10^8 \msun$, all models assume $a=0.95$. (a) Electron temperature as a function of the bolometric Eddington ratio. The (orange) circles  are for $\beta=9$ and $\delta=10^{-3}$ with $\dot m = 0.1$, 0.6 and 1 from left to right; the (red) stars are for $\beta=1$ and $\delta=10^{-3}$ with $\dot m = 0.1$, 0.3, 0.45 and 0.6;
the (blue) squares are for $\beta=1$ and $\delta=0.5$ with $\dot m = 0.033$, 0.1 and 0.3;
the (black) crosses are for models with an outflow, o4 (lower) and o5 (upper);
the right (magenta) open circle is for $\beta=9$, $\dot m = 0.1$ and  $\delta=0.5$;  
the (green) triangles are for magnetically-dominated models,
the triangles up are for $\beta=0.3$ and $\delta=10^{-3}$ with $\dot m = 0.1$, 0.3, 0.5; triangles down are for $\beta=0.43$ and $\delta=0.5$ with $\dot m = 0.033$, 0.1, 0.3. The solid, dashed and dotted lines show the effect of the change of $\dot m$, $\delta$ and $\beta$, respectively.  (b) $T_{\rm e}^{\rm PS}$ as a function of $\tau^{\rm PS}$; all symbols have the same meaning as in (a).
The open star shows parameters of the lower $T_{\rm e}^{\rm PS}$ fit to model s8, discussed in Section \ref{sect:te}.
(c) $T_{\rm e}^{\rm PS}$ as a function of $\tau^{\rm PS}$; the stars are for $\beta=1$ and $\delta=10^{-3}$ with $\dot m = 0.1$, 0.3, 0.5; other symbols have the same meaning as in (a) except for $M =2 \times 10^8 \msun$. For GX 339-4 and Cyg X-1 we use  
$\tau^{\rm PS}$ from the sphere fits  reduced by a factor of 1.5 and for NGC 5548 by a factor of 2 (see Section \ref{sect:compps}); for other objects we use the slab fits. 
}
\label{fig:comppsparam} 
\end{figure*}

In all models, the increase of electron heating with increasing $\dot m$ is outweighed by the enhancement of the cooling rate due to the increase of $\tau$ ($\propto \dot m$). Then, the electron temperature decreases with increasing $\dot m$, see Fig.\ 2b, which is also reflected in the dependence of $T_{\rm e}^{\rm PS}$ on $\dot m$ (see Fig.\ \ref{fig:comppsparam}).

If direct heating of electrons dominates, $L$ increases linearly with $\dot m$, i.e.\ $\eta$ is constant (see also fig.\ 1 in Xie \& Yuan 2012). Flows with $\delta = 0.5$ have large efficiencies, $\eta \ga 0.1$, for large $a$. Note also that for $\delta=0.5$ the values of $\eta$ are comparable with those of a standard, Keplerian disc with the same $a$, $\eta_{\rm NT}$, as given by the Novikov \& Thorne (1973) model; namely, $\eta \simeq 0.3-0.8 \eta_{\rm NT}$, depending on $a$ and $\beta$.  

The Coulomb rate increases much faster than the compressive heating rate, therefore, in small-$\delta$ models a transition between the compression- and Coulomb-dominated heating occurs at some  $\dot m$ (e.g.\ in our $a=0.95$ models, at $\dot m \simeq 0.1$ for $\beta=1$ and at $\dot m \simeq 0.3$ for $\beta=9$). For 
the total Coulomb rate we find $\Lambda_{\rm ie,tot} \propto \dot m^{2.5}$; the obvious scaling with the square of density is further enhanced by the decrease of $T_{\rm e}$  with increasing  $\dot m$ (increasing the difference between $T_{\rm i}$ and $T_{\rm e}$). Then, in flows dominated by  the Coulomb heating $\eta \propto \dot m^{1.5}$. The total compressive heating rate $Q_{\rm compr,tot} \propto \dot m^{0.6}$, where the increase of density is partially compensated by the decrease of $T_{\rm e}$. In models with heating dominated by the compression work, i.e.\ those with high $\beta$, small $\delta$ and small $\dot m$, we find $\eta \simeq 0.004$ regardless of other parameters.

For small $\delta$, we can expect a much stronger dependence of  $\eta$ on $\dot m$ for $\dot m > \dot m_{\rm up}$, where cooling of ions should lead to the decrease of $H/R$ and the increase of density. Our models s10 and s11, with $\dot m \simeq \dot m_{\rm up}$ for $\beta=1$,  have moderate $\eta \simeq 0.02$, $L \simeq 0.01 \ledd$ and $\Lambda_{\rm ie,tot} < 0.1 Q_{\rm i,tot}^+$. Then, around the critical accretion rate, $\dot m_{\rm crit}$, where $\Lambda_{\rm ie,tot} \simeq Q_{\rm i,tot}^+$, we can expect an order of magnitude larger $\eta$. Similarly, for $\delta=0.5$ we can expect a comparable contribution of the direct and Coulomb heating of electrons around $\dot m_{\rm crit}$, yielding $\eta$ twice larger than at $\dot m < \dot m_{\rm up}$. Such a rapid increase of $\eta$, by a factor of 2 for $\delta=0.5$ and by almost an order of magnitude for $\delta=10^{-3}$, related to the rapidly increasing $\Lambda_{\rm ie}$ at $\dot m$ approaching the critical value, is clearly seen in fig.\ 1 of Xie \& Yuan (2012).

The increase of $\dot m$ results in a steep $\Gamma$--$\lambda_{2-10}$ relation, with an increase of $\lambda_{2-10}$ by a factor of 10 corresponding to hardening by $\Delta \Gamma \simeq -(0.4$--0.5), see Fig.\ \ref{fig:evol}a, although we note hints for a flatter relation at small $\dot m$ in models dominated by compressive heating. The  relation is rather weakly dependent on parameters, for a given $M$ it depends mostly on $\delta$ (see Section \ref{sect:delta}).
Even a crude parametrisation of involved processes in Gardner \& Done (2013) gives a similarly steep relation, however, we note quite large quantitative differences, in particular, their relation is shifted to lower luminosities (by a factor of $\sim 5$).

\subsection{Outflow}
\label{sect:outfl}

As can be seen in Fig.\ \ref{fig:flow}a, presence of an outflow leads to a significant decrease of the $H/R$ ratio. As a result, in this class of models the flow resembles a slab most closely. The outflow also strongly reduces   $\tau_z$ in the inner region, e.g.\ in models o2 and o4 we get a 
constant $\tau_z \simeq 0.05$ at $r<100$. The reduced $\tau_z$ results in a strong increase of $T_{\rm e}$ toward the event horizon, unlike models with $s=0$ where a decrease or at most a weak increase of $T_{\rm e}$  toward small $r$ occurs, see Fig.\ \ref{fig:temp}. The steep  $T_{\rm e}$ profile for $s>0$ leads to a specific radial distribution of the radiation field components, with a strongly centrally concentrated synchrotron emission (bulk of which comes from $r \la 2$) and a much flatter Comptonized component. As a result,  the global nature of the Compton process is more important here than in models with $s=0$. Namely, for $s=0.3$ the flux of synchrotron photons produced at $r_{\rm max}$ ($\simeq 10$, see definition in Section \ref{sect:compps}) is much smaller than the flux of synchrotron photons transferred from $r \la 2$, i.e.\ electrons giving the dominating contribution to the observed X-ray spectra cool on seed photons produced in a region of much larger $T_{\rm e}$.

However, the major effect of the outflow concerns the slab-like geometry in models with $\beta=9$. For $\beta=9$, models with an outflow predict much smaller $T_{\rm e}^{\rm PS}$ than models with $s=0$. This again shows that changes of the geometrical shape of the flow play the dominating role for the magnitude of $T_{\rm e}$. For $\beta=1$, presence of an outflow only weakly affects the value of $T_{\rm e}^{\rm PS}$. Also regardless of the value of $\beta$, we find that the outflow models follow the same $\Gamma$--$\lambda_{2-10}$ relation  as models without an outflow. 

The most prominent effect of the outward transfer of seed photons from the inner, large-$T_{\rm e}$ region, occurs  at $r \sim 100$,  where $\tau_r$ is relatively high for $s>0$, and $T_{\rm e}$ is reduced by a factor of $\sim 3$ as compared to the initial solution with local prescription for Compton cooling (the effect illustrated in  figure 3a in Xie et al.\ 2010). As a result, the Coulomb rate is significantly larger in the global-Compton outflow model mostly at $r \simeq 50$--300, where also the $\Lambda_{\rm ie}/Q_{\rm i}^+$ ratio is several times larger in the $s=0.3$ than in the $s=0$  models with the same $L$.

The efficiency of our outflow model, $\eta \simeq 0.01$, is in approximate agreement with Xie \& Yuan (2012), who find a constant $\eta \simeq 0.003$ (scaled to our definition) in the same range of $\dot m_{\rm out}$; the difference between these magnitudes of $\eta$ can be understood as the result of a smaller dissipation rate, by a factor of several as compared to our model with $a=0.95$, and the neglect of GR transfer effects (which reduce the escaping flux by up to 50 per cent), in the nonrelativistic model in Xie \& Yuan (2012).

\begin{figure} 
\centerline{\includegraphics[height=7.8cm]{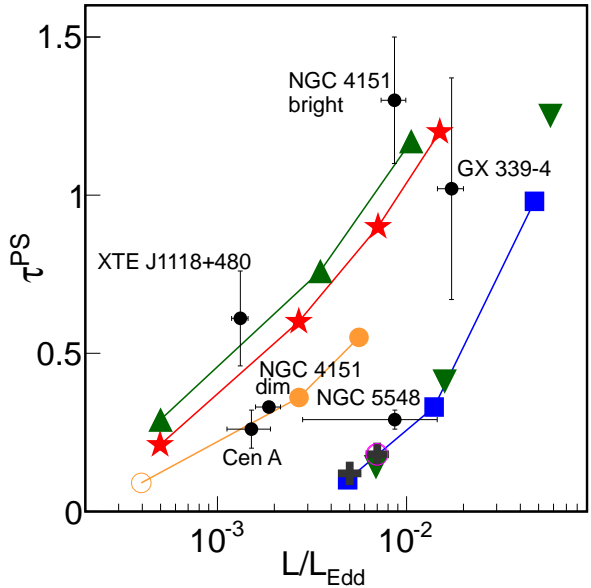}}
\caption{Model points show $\tau^{\rm PS}$ as a function of the bolometric Eddington ratio for our models with $M = 10\, \msun$ (differences for the models with $M = 2 \times 10^8 \msun$ are insignificant). All symbols have the same meaning as in Fig.\ \ref{fig:comppsparam}. For  GX 339-4 and XTE J1118+480 we use bolometric luminosities estimated by Miyakawa et al.\ (2008) and Frontera et al.\ (2003). For AGNs we use the approximation $(L/\ledd)=7 \lambda_{2-10}$. If only sphere fits for the observed data are available, we use $\tau^{\rm PS}$ reduced as in Fig.\ \ref{fig:comppsparam}.
}
\label{fig:taulum} 
\end{figure}	

\subsection{Black hole mass}

Models with $M =2 \times 10^8\, \msun $ in general predict smaller $\Gamma$ and higher $T_{\rm e}^{\rm PS}$ than models with $M = 10\, \msun $, see Figs\ \ref{fig:1}b and \ref{fig:evol}a, which property results from the dependence of the thermal synchrotron power on $M$ (see also  Wardzi\'nski \& Zdziarski 2000).
In the considered range of $\dot m$, with a negligible Coulomb cooling of  protons, the hydrodynamical solution is essentially $M$-invariant, i.e.\ $v^r$, $v^\phi$,  $H$, $T_p$, $n$ and $Q_{\rm vis}/L_{\rm edd}$ are the same functions of $r$ independent of the value of $M$. However, for electrons a non-trivial dependence on $M$ occurs due to  the scaling of the thermal synchrotron power, $L_{\rm synch,th}/L_{\rm edd} \propto M^{-1/2}$ (Mahadevan 1997).
This, with the Eddington ratio of the total heating being roughly independent of $M$ (see below), implies that $T_{\rm e}$ increases with $M$. E.g.\, for  $\delta=10^{-3}$ and $\dot m=0.1$, in models with $M = 2 \times 10^8 \, M_{\odot}$ the temperature is larger by 60--100 keV (for $\beta=1$) and by 150--200 keV (for $\beta=9$) than in models with $M = 10\, M_{\odot}$, see Fig.\ 2. Corresponding to these changes of $T_{\rm e}$, $Q_{\rm compr,tot}/L_{\rm Edd}$ is smaller, while $\Lambda_{\rm ie,tot}/L_{\rm Edd}$ is larger for $M = 10\, M_{\odot}$, both  by $\sim 30$ per cent, so that 
the Eddington ratio of the total heating remains independent of $M$.

\subsection{Viscosity parameter}
\label{sect:viscosity}

In agreement with previous studies, e.g.\ Narayan \& Yi (1995), Esin et al.\ (1997), YZ04, we find that a large $\alpha$ is required to explain luminous hot flows, in particular $\alpha \sim 0.3$ (a rough estimate based on $\Lambda_{\rm ie}/Q_{\rm i}^+$) is needed for $L \ga 0.1 \ledd$ (see Section \ref{sect:turn}).
In our hydrodynamical model we find $n \propto \alpha^{-1}$, which should lead to $\dot m_{\rm crit} \propto \alpha^2$ as in the self-similar model of Narayan \& Yi (1995).

Therefore, we do not make a detailed study of the spectral dependence on $\alpha$. However, we have checked that the change of $\alpha$ does not affect spectral properties (provided that the Coulomb cooling of ions remains negligible) so our results should remain valid even if  $\alpha$  is  $\dot m$-dependent.
E.g.\ model v1, with $\alpha=0.1$, follows the same  $\Gamma$--$\lambda_{2-10}$ (see Fig.\ \ref{fig:evol}a) as well as $T_{\rm e}^{\rm PS}$--$\lambda_{2-10}$ relations as the corresponding model with $\alpha=0.3$. On the other hand, the radiative efficiency depends on $\alpha$ and we find that for small $\delta$ it does not follow a simple $\eta \propto \alpha^{-2}$ relation (which could be inferred from the $n \propto \alpha^{-1}$ dependence) mostly due to smaller $T_{\rm i}$, and hence comparatively weaker Coulomb heating, for smaller $\alpha$; the effect is seen in comparison of model v1 (with $\dot m=0.1$) with s9 (with $\alpha=0.3$ and $\dot m=0.3$, yielding the same density as in v1).

\section{Comparison with observations}
\label{sect:obs}

In this Section we compare our results with the X-ray spectral data for three  thoroughly studied transient black-hole binaries and we find a rough agreement at $L < 0.01 \ledd$. We also attempt to make a similar, precise comparison of our results with AGNs observed at $L < 0.01 \ledd$. Here the observational grounds are more uncertain and the derived parameters of intrinsic X-ray emission depend on approach to spectral modelling. Nevertheless, the comparison with several best-studied AGNs provides a compelling evidence for a systematic disagreement between the model and data.

We  study the hardness vs Eddington ratio evolution using $L_{2-10}$. The bolometric $L$ is often used for such studies, however, $L_{2-10}$ is a directly measurable quantity, whereas estimates of $L$ assume some $L/L_{2-10}$  scaling (typically $\sim 10$), which approach may introduce inaccuracies by a factor of $\sim 2$ (the amount of changes of  $L/L_{2-10}$ in our models). We find $L_{2-10}$ of the intrinsic power-law component using the values of power-law  index and normalization collected from references listed below. When available, uncertainties on $M$ and distance, $d$, are used to plot error bars on $\lambda_{2-10}$.

\begin{figure*} 
\centerline{\includegraphics[height=10cm]{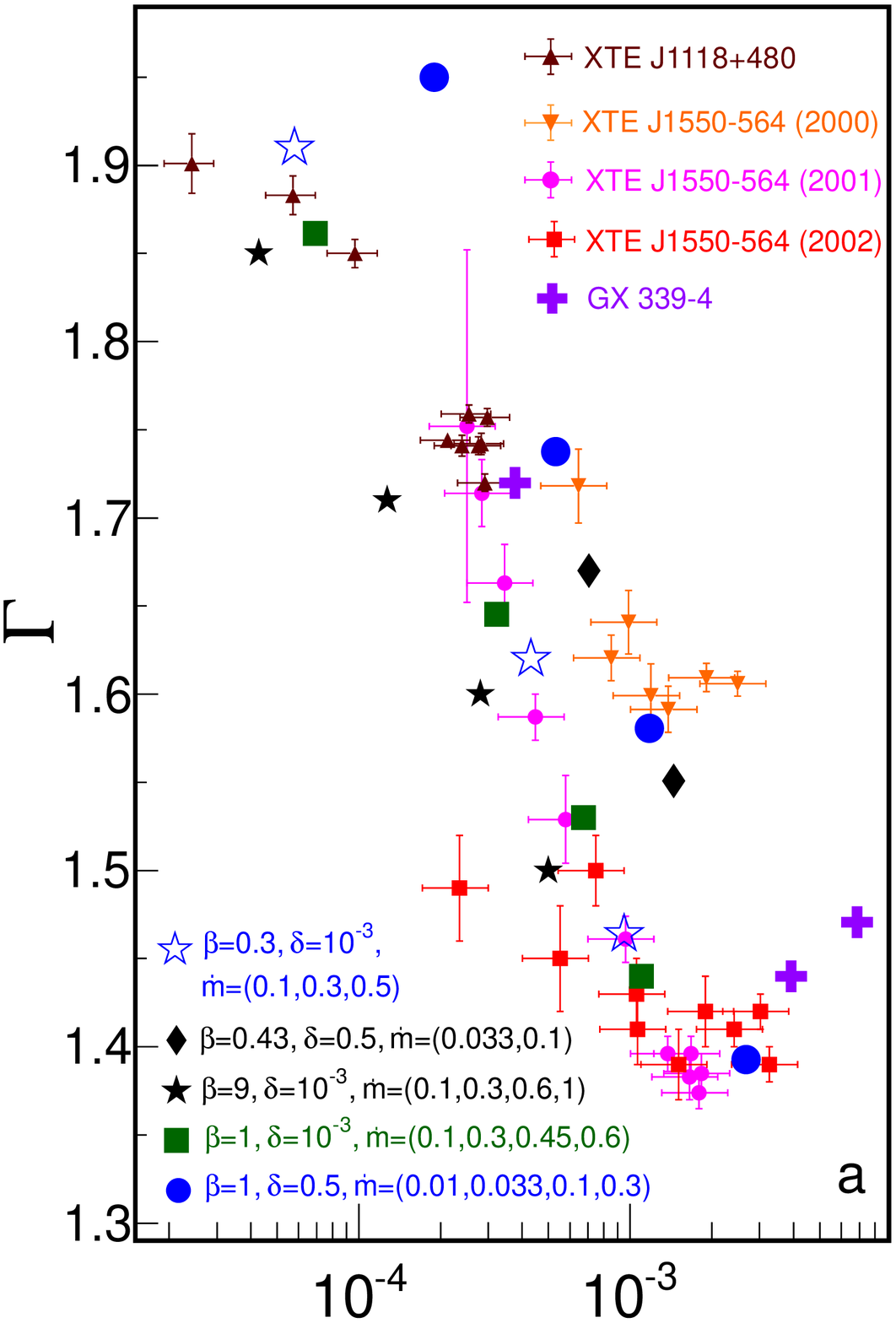}
\includegraphics[height=10cm]{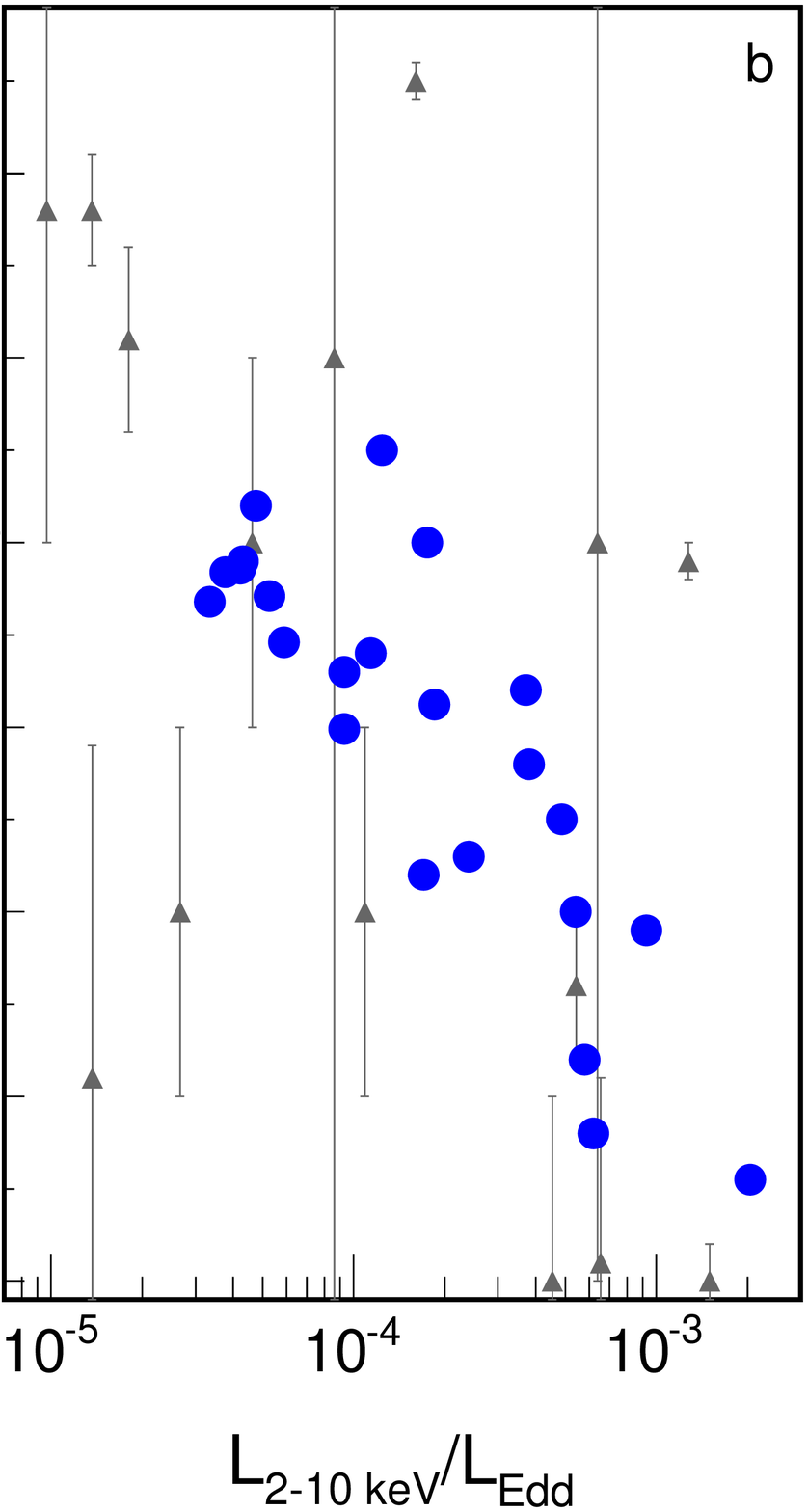}
\includegraphics[height=10cm]{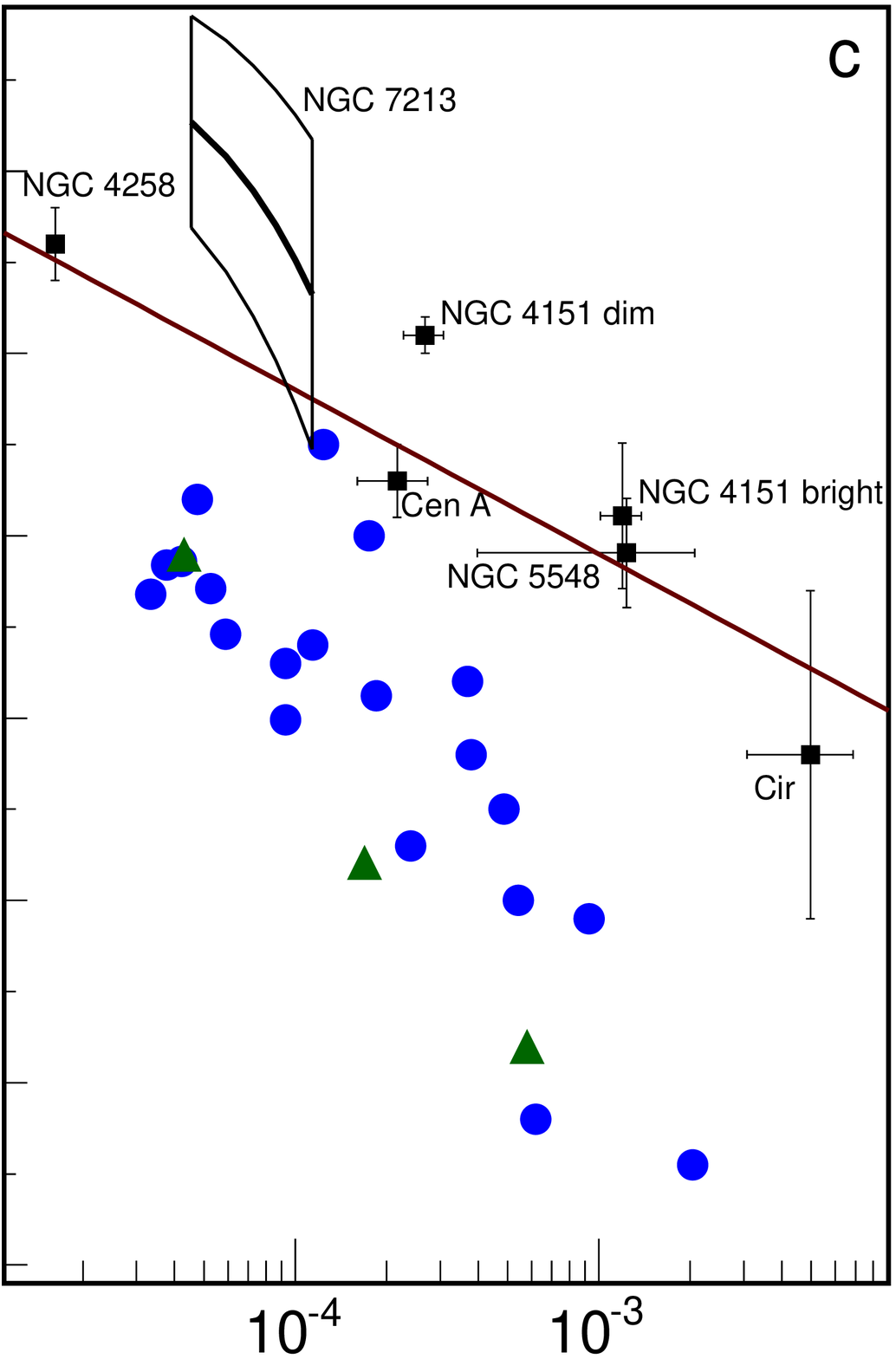}}
\caption{Photon spectral index as a function of the 2--10 keV Eddington ratio, predictions of the hot flow model are compared with the observational data for black hole binaries (a) and for low-luminosity AGNs (b,c). Panel (a) shows the data sets for the decay of 3 outbursts of XTE J1550-564, the decay of  XTE J1118+480 and the rising phase of GX 339-4; all model points are for $M = 10 \, \msun$, $a=0.95$ and other parameters  as labelled in the figure. Panel (b) shows the data for low-luminosity Seyferts from Gu \& Cao (2009) and panel (c) shows parameters of the intrinsic X-ray emission for well-studied AGNs (see text), as labelled. The model points in panels (b) and (c) include all the model results computed in this work (i.e.\ those shown in Fig.\ \ref{fig:evol} or in Tables \ref{tab:tab1} and \ref{tab:tab2}) for  $M =2 \times 10^8\, \msun $. Triangles in (c) are for $\beta=0.3$ and $\delta=10^{-3}$.  The solid line in (c) shows the fit from Gu \& Cao (2009).
}
\label{fig:observ} 
\end{figure*}	

\subsection{The hard state of black-hole transients}
\label{sect:hardstate}

\subsubsection{$\Gamma$--$\lambda_{2-10}$ correlation for a plain hot-flow emission}
\label{sect:outbursts}

Transient black hole binaries show hysteresis in the dependence of the spectral shape on $L$ (e.g.\ Zdziarski \& Gierli\'nski 2004) which, in terms of the scenario involving transition between a cold disk and a hot flow, implies that the disc appears at higher $L$ during the rise and disappears at lower $L$ during the decay of an outburst. 

Changes of the truncation radius of the cold disc, $r_{\rm tr}$, with the change of $L$  have been a subject of intensive studies and discussion over the recent years. Using the amount of relativistic distortion of the reflected component, Plant et al.\ (2013) find $r_{\rm tr} \ga 600$ at  $\lambda_{2-10} \simeq 4 \times 10^{-4}$, $r_{\rm tr} \ga 400$ at $\lambda_{2-10} \simeq 4 \times 10^{-3}$, $r_{\rm tr} \sim 200$ at $\lambda_{2-10} \simeq 7 \times 10^{-3}$ and $r_{\rm tr} \sim 100$  at $\lambda_{2-10} \simeq 0.013 $ during the {\it rising} phase of GX 339-4, indicating that the inner edge of the thin disk moves slowly to small radii and even at very large $L \sim 0.1 \ledd$ the hot flow is not affected by external irradiation (we have checked that irradiation of the inner flow is negligible for $r_{\rm tr} \ga 100$). During the {\it decay}, Plant et al.\ (2013) find a much smaller $r_{\rm tr} \sim 10$  at $\lambda_{2-10} \simeq 3 \times 10^{-3}$ and the spectral index, $\Gamma \simeq 1.8$, being much softer than that during the rise, $\Gamma \simeq 1.4$, at the same $\lambda_{2-10}$. The above determination of $r_{\rm tr}$ relies on the assumption that the X-ray source is not located at a distance larger than $r_{\rm tr}$ (cf.\ Fabian et al.\ 2014), i.e.\ - considering the above values of $r_{\rm tr}$ - that the X-ray emission is not dominated by a jet or the base of a jet. This is then related to the problem of the flow or jet dominating the X-ray emission; we regard the former to be more likely basing on the rough agreement of GX 339-4 both with other transients and with prediction of the hot flow model, see below  (see also references in Section 1).

An independent assessment of  $r_{\rm tr}$ involves  modelling of the disc thermal emission. Cabanac et al.\ (2009) find that  the cold disc extends close to the black hole at the bolometric $L \ga 10^{-2} \ledd$ and it starts to recede at $L \sim 10^{-3}-10^{-2} \ledd$. Their study involves mostly the decay phases, hence these results are consistent with the above assessments of Plant et al.\ (2013).

In Fig.\ \ref{fig:observ}a we compare our results with two black-hole binaries, XTE J1118+480 and XTE J1550-564, with precisely determined black hole masses.
We use four  {\it RXTE} data sets obtained during (1) the end of the decay of XTE J1118+480, and the decay phases of 3 outbursts of  XTE J1550-564: (2) the big outburst in 2000, (3) the  mini-outburst of in 2001 and (4) the mini-outburst in 2002. Datasets (1-3) are from Kalemci (2000) and (4) is from Belloni et al.\ (2002). For XTE J1118+480 we adopt the distance of $d = 1.7 \pm 0.1$ kpc and the black hole mass $6.9M_\odot \le M \le 8.2 \, M_\odot$ from Khargharia et al.\ (2013) and  for XTE J1550-564, $M = 9.1 \pm 0.6 \, M_\odot$  and $d = 4.38^{+0.58}_{-0.41}$ kpc, from Orosz et al.\ (2011). We also use (5) the data for the outburst rise of GX 339-4 from  Plant et al.\ (2013), with $\lambda_{2-10}$ corresponding to   $r_{\rm tr} \ga 200$ (see above); we adopt $M = 8\, \msun $ and  $d=8$ kpc, which values are uncertain (but likely, cf.\ Zdziarski et al.\ 2004) and for GX 339-4 we do not plot error bars on the Eddington ratio. 
 
The 2000  outburst of XTE J1550-564 showed transition into the soft (disc-dominated) state, see e.g.\ Russell et al.\ (2010), whereas during the weaker 2001 and 2002 outbursts it was only observed in the hard state (e.g.\ Belloni et al.\ 2002) and most likely the transition luminosity  was not reached. Furthermore, during the 2001 and 2002 declines the spectrum is much flatter than during the 2000 decline, in turn, as seen in Fig.\ \ref{fig:observ}a, the $\Gamma$--$\lambda_{2-10}$ relation in these weaker-outbursts declines agrees with that observed  during the rise in GX 339-4. This, with the findings on  $r_{\rm tr}$ noted above, indicates that during the rising phases (dataset 5) as well as the whole weak outbursts, when an inner cold disc is not built (datasets 3 and 4), we observe a plain emission of the hot flow, not affected by irradiation from the outer disc.

In the intensity-hardness diagrams, the decline branch typically merges with the rising one at $\lambda_{2-10} \simeq (2-3) \times 10^{-4}$ (see e.g. fig.\ 2 in Miyakawa et al.\ 2008). This indicates that at such luminosities the cold disc recedes beyond at least several tens of $R_{\rm g}$ and such low-$\lambda$ observations can be compared with our results (neglecting external irradiation). The late decay evolution, with $\lambda_{2-10} \la  2 \times 10^{-4}$, is represented by the data set (1). Strongly supporting the absence of the inner disc at such $\lambda_{2-10}$  during the  decay of XTE J1118+480, 
Esin et al.\ (2001) find $r_{\rm tr} \sim 100$ (possibly underestimated by  a factor of a few, Gierli\'nski et al.\ 2008) by spectral fitting. Assessments of much lower   $r_{\rm tr}$ at $\lambda_{2-10} \simeq 10^{-4}$ are addressed in Section  \ref{sect:bump}.

As we see in Fig.\ \ref{fig:observ}a, the datasets (1), (3), (4) and (5), most likely representing the plain hot-flow emission, are in good agreement with the model prediction for $\beta=1$ (or lower) and $\delta = 10^{-3}$ at $\lambda_{2-10} \la 10^{-3}$, except for the two lowest-$L$ points in (4). Models with $\delta = 0.5$ tend to predict slightly too soft spectra and the discrepancy increases with decreasing $\lambda_{2-10}$. The evolution toward very hard spectra, with $\Gamma \simeq 1.5$ at $\lambda_{2-10} \simeq 2 \times 10^{-4}$, as in the data set (4), cannot be explained by our hot flow model.  

During the decline of the strong outburst of XTE J1550-564 in 2000, the hot flow should be strongly irradiated  by a weakly-recessed disc. Note that our data set (2) includes only the more luminous part of this decline, when the X-ray radiation was dominated by thermal Comptonization (cf.\ Russell et al.\ 2010). By comparison with other datasets, it illustrates the difference between the $\Gamma$--$\lambda_{2-10}$ relation for hot-flow emission affected and not affected by external irradiation. We  note that we regard the agreement of the model for $\delta=0.5$ with the dataset (2) as incidental  because the model does not include the input of soft photons from the cold disc, which should be strong during this observation.

\subsubsection{Electron temperature} 
\label{sect:te}

A strong anticorrelation between the cut-off energy and luminosity is observed above $0.01 \ledd$, however, at lower $L$ the cut-off is  typically  not well constrained due to a poor photon statistics (see e.g.\ Miyakawa et al.\ 2008, Yamaoka et al.\ 2005). The fits with the COMPPS model
in Miyakawa et al.\ (2008) show the decrease of $kT_{\rm e}^{\rm PS}$ from $\sim 90$ keV at $L \simeq 0.01 L_{\rm Edd}$ to $\sim 50$ keV  at $L \simeq 0.1 L_{\rm Edd}$ in GX 339-4. This behaviour is qualitatively consistent with predictions of the hot-flow model for  changes of $L$ resulting from the change of $\dot m$, see Fig.\ \ref{fig:comppsparam}a.  However, in this range of $L$ we can directly compare the measured $kT_{\rm e}^{\rm PS}$ only with predictions of our model for $\delta=0.5$. Our largest-$L$ solution for  $\delta=10^{-3}$, with a similar luminosity to the lowest-$L$ fit (shown in Fig.\ \ref{fig:comppsparam}a) from Miyakawa et al.\ (2008), is at the upper limit of the measured $kT_{\rm e}^{\rm PS}$ (this model temperature may be slightly overestimated, see Section \ref{sec:flow}).
We note that extrapolation of the small-$\delta$ branch to $L > 0.01 L_{\rm Edd}$ is very uncertain, because a  quantitative assessment of effects related to significant cooling of ions is not   feasible without explicit computations.

Notable measurements of small electron temperatures, $kT_{\rm e}^{\rm PS} \simeq 100$ keV, at low luminosities of $L \sim 0.001 \ledd$, concern XTE J1118+480 (Frontera et al.\ 2003) and 1E1740.7-2942 (Natalucci et al.\ 2014; assuming $d=8.5$ kpc). Whereas the latter is a rather poorly understood black-hole candidate,  XTE J1118+480 is the standard source motivating hot-flow models and explanation of the measured value of $kT_{\rm e}^{\rm PS}$ strongly challenges the model.
 Parameters of the fit from Frontera et al.\ (2003)\footnote{We use ToO3 from table 4 (where $kT_{\rm seed} = 10$ eV is assumed to approximate the thermal synchrotron input)  which has a better fit quality than ToO1 and smaller uncertainties on fitted parameters than ToO2. Note also that there seems to be an inconsistency in their definition of model parameters, possibly their $\tau$ denotes a half-thickness for ToO1 and a full-thickness for ToO3, otherwise the given $kT=68$ and 112 keV, respectively, with the same $\tau \approx 0.6$, would produce a large change of the slope, $\Delta \Gamma \simeq 0.3$, contrary to what is shown in figures 3 and 4 in Frontera et al.\ (2003).} are shown in Fig.\ \ref{fig:comppsparam}. Formally, it requires  a super-equipartition magnetic field and even $\beta=0.3$ gives $kT_{\rm e}^{\rm PS} \simeq 140$ keV, i.e.\ slightly above the fitted range.

\begin{figure} 
\centerline{\includegraphics[width=8cm]{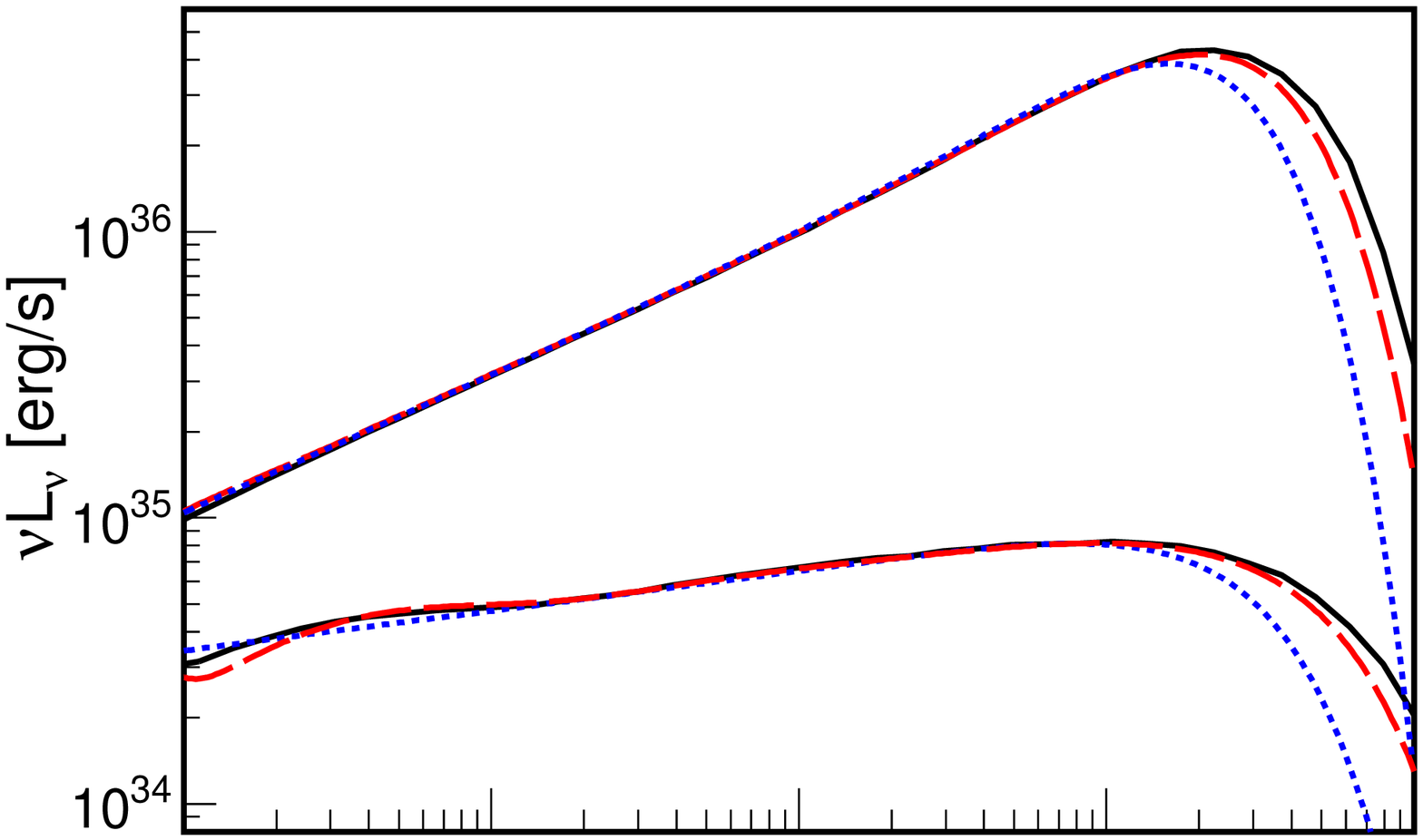}}
\centerline{\includegraphics[width=7.94cm]{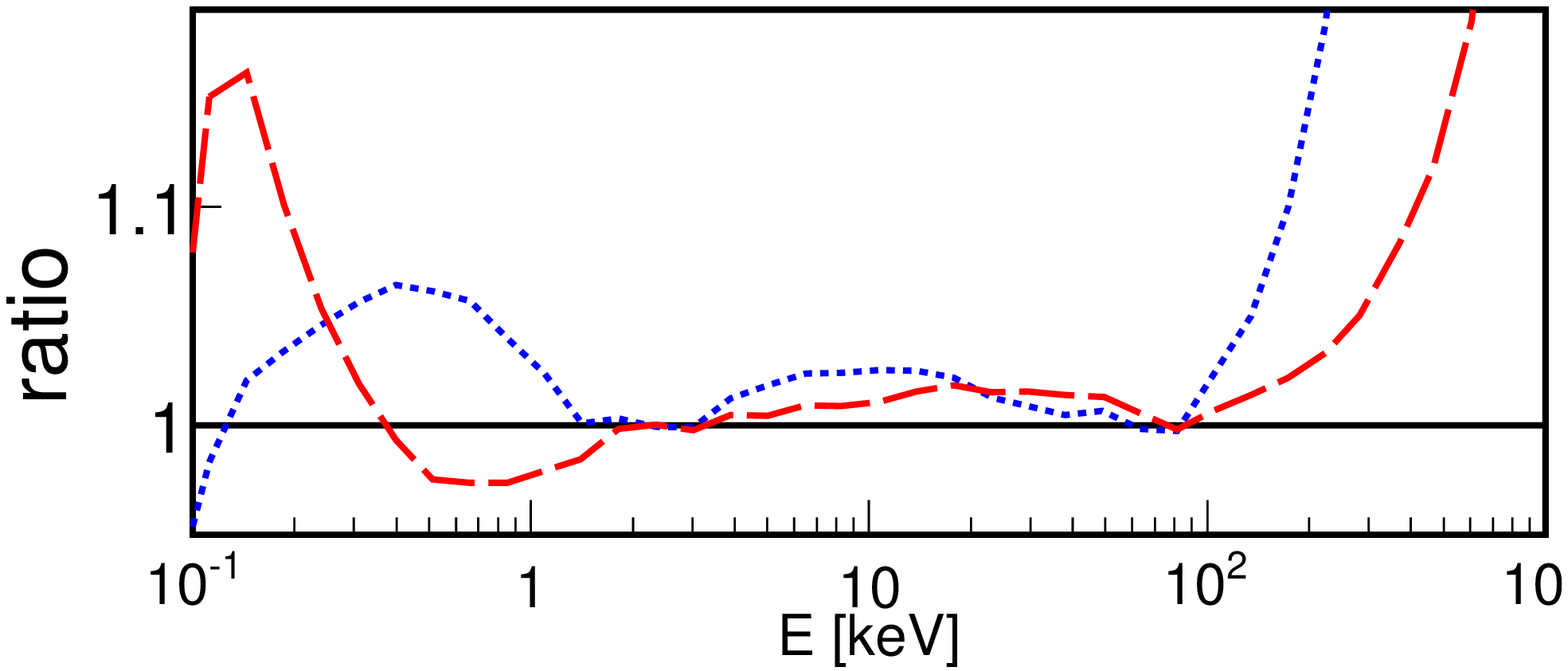}}
\caption{Top panel: the upper solid spectrum is for model s14, the dashed and dotted spectra show its COMPPS fits with ($kT_{\rm e}^{\rm PS}$, $\tau^{\rm PS}$)=(110 keV, 1.17) and (80 keV, 1.55); the lower solid spectrum is for model s8,  the dashed and dotted spectra show COMPPS fits with 
(206 keV, 0.21) and (120 keV, 0.49), respectively. Residuals to the fits of the latter (s8) are shown in the bottom panel. The COMPPS models assume $kT_{\rm seed} = 10$ eV.
}
\label{fig:ecut} 
\end{figure}

However, we should make a remark here on the precision of measurements of $T_{\rm e}$ in this range of parameters. As shown in Fig.\ \ref{fig:ecut},
a significant change, by $\sim 90$ keV, of $kT_{\rm e}^{\rm PS}$ for the fits of model s8 (approximately relevant for XTE J1118+480)
does not change $E_{\rm cut}$ ($\simeq 100$ keV, as given by  the $\nu F_\nu$ maximum). In turn, a much smaller change of $kT_{\rm e}^{\rm PS}$, by 30 keV,  for the fits of model s14  yields a significant change of $E_{\rm cut}$ by 40 keV. Then, for harder spectra 
(as in model s14) $T_{\rm e}$ is constrained just by the  cut-off position, whereas for softer (as in s8) high-quality data above  $E_{\rm cut}$  are needed for a precise constraint on $T_{\rm e}$. The 100--200 keV flux of our higher and lower $T_{\rm e}^{\rm PS}$ fits for model s8 differs by only 5 per cent.
Frontera et al.\ (2003) use only the data for photon energies lower than 200 keV and it seems highly unlikely that such a small difference could be constrained with the quality of data in their high energy bins, as shown e.g.\ in their figure 3. 
More likely their constraint on $T_{\rm e}^{\rm PS}$ results from the difference of spectral shapes in the soft X-ray range, where the COMPPS model rather poorly approximates hot-flow spectra for small $\tau$ and the fits should favour a lower $T_{\rm e}^{\rm PS}$ (and higher $\tau^{\rm PS}$). Such higher $\tau^{\rm PS}$ spectra are closer to a power-law and give residuals which can be approximated by a blackbody component (see also Section \ref{sect:bump}), while for higher $T_{\rm e}^{\rm PS}$ the residuals have a more complex shape, see Fig.\ \ref{fig:ecut}. A similar effect is noted by Lubi\'nski et al.\ (2010) in their discussion of constraints on $T_{\rm e}^{\rm PS}$ for the dim state of NGC 4151.

Parameters of the lower $T_{\rm e}^{\rm PS}$ fit to model s8 are shown by an open star in Fig.\ \ref{fig:comppsparam}b and taking into account the above we suspect that the observation of  XTE J1118+480 may be consistent with hot flow models for $\beta \le 1$ and $\delta=10^{-3}$ if a more accurate description of spectral shape is taken into account. On the other hand, models with large $\beta$ or $\delta$   strongly overpredict the electron temperature (by hundreds of keV) and are ruled out by the observed cut off.

At $L \sim 0.01 \ledd$ all models with $\delta=0.5$, including the magnetically-dominated model s4, predict $kT_{\rm e}^{\rm PS} \ga 200$ keV, strongly exceeding $kT_{\rm e}^{\rm PS} \simeq 50$--90 keV measured in GX 339-4  in this range of $L$, which again disfavours models with strong direct heating of electrons. 
In Fig.\ \ref{fig:comppsparam}b we show also the fits to Cyg X-1 from Gierli\'nski et al.\ (1997; we use single-temperature sphere fits). As we note in Section  \ref{sect:summary}, this object is likely not relevant for the comparison with our model, yet we note that the fitted $\tau$ and $T_{\rm e}$ can be explained by hot-flow models with small $\delta$ and $\beta$ parameters. 

An attractive explanation of low electron temperatures could involve an additional, internal source of seed photons, e.g.\  the nonthermal synchrotron radiation 
is an obvious candidate within the framework of hybrid model, see Poutanen \& Veledina (2014). We note, however, that the presence of such a source, much stronger than the thermal synchrotron radiation, seems to be ruled out in black hole transients, at least for $\lambda_{2-10} \la 10^{-3}$. We have checked that the increased Compton cooling  resulting from such an extra source in general leads to the increase of both $\Gamma$ and $\lambda_{2-10}$, i.e.\ the model points move toward the top-right part of the $\lambda_{2-10}$--$\Gamma$ diagram, resulting in a discrepancy with the observed $\lambda_{2-10}$--$\Gamma$ correlation. This effect is seen also in figure 12 in Veledina, Vurm \& Poutanen (2011), where their hybrid model gives a much flatter $\Gamma$--$\lambda_{2-10}$ relation, with larger values of $\Gamma$, than predicted by our purely-thermal model at $\lambda_{2-10}<10^{-3}$.

As we see in Fig.\ \ref{fig:taulum}, models with $\delta=0.5$ predict  $\tau^{\rm PS}$ much lower than required by the spectral fits of XTE J1118+480 and GX 339-4 and we note that this theoretical constraint is essentially independent of the presence of a stronger source of seed photons.

Although the value of $E_{\rm cut}$ provides a strong constraint on the hot flow model, this issue has been addressed in only a few previous studies. Esin et al.\ (1998), using an approximate treatment of the global Compton cooling as well as taking into account the change of the adiabatic index with $\beta$, find $T_{\rm e} \simeq 130$--110 keV at $r=20$ in the model with $\beta=1$ and $\delta=10^{-3}$, see their figure 2. Those values of $T_{\rm e}$ seem to be in an approximate agreement with our model, although Esin et al.\ (1998) present $T_{\rm e}$ only for $\dot m \ga 1$ (i.e.\ around the critical accretion rate, cf.\ figure 6b in Esin et al.\ 1997); our model s11 with the same $\beta$ and $\delta$ and slightly smaller $L$  has $T_{\rm e} (r=20) \approx 140$ keV.
On the other hand, we note a large discrepancy between our results and Esin  et al.\ (2001), who use the model  refined 
by the GR description. The model spectra presented in their figure 1 show cut-off energies much lower than found in our models with the same $L$, e.g.\ for $\beta=1$ and $\delta=10^{-3}$ we find an over a factor of 3 larger $E_{\rm cut}$. 

Yuan \& Zdziarski (2004) find that their standard hot-flow solutions with $\delta=10^{-3}$ are characterized by   $T_{\rm e} \sim 150$--200 keV at $L \simeq (0.03$--$0.06) L_{\rm Edd}$. We note that those results were obtained with the nonrelativistic, local Compton model, so their accuracy is somewhat uncertain, furthermore,  $\beta=9$ was assumed in their computations for which the model  predicts the largest $T_{\rm e}$.

\subsubsection{Departures from a power-law at small $\lambda_{2-10}$} 
\label{sect:bump}

The spectra in models with $\dot m \sim 0.1$ exhibit 
individual scattering bumps, more pronounced for larger $\beta$ and $\delta$ and smaller $\dot m$ (i.e.\ larger $T_{\rm e}$  and/or smaller $\tau$). For $M \sim 10 \, \msun $, with (thermal synchrotron) seed photons at $\sim 10$--100 eV, the first scattering bump occurs in soft X-ray range. In some spectra, e.g.\ the bottom one in Fig.\ 4a corresponding to $\lambda_{2-10} \sim 5 \times 10^{-5}$, the second scattering bump is also clear above $\sim 10$ keV. Interestingly, spectra with such a concave shape are reported at relevant $\lambda_{2-10}$, e.g.\ the faint spectrum of GX 339-4  with a strong turn up above $\sim 7$ keV in Wardzi\'nski et al.\ (2002; see their figure 4, note that their interpretation is incorrect - the suggested strong contribution of bremsstrahlung can be expected at $L$ by at least  two orders of magnitude smaller).

\begin{figure} 
\centerline{\includegraphics[width=8cm]{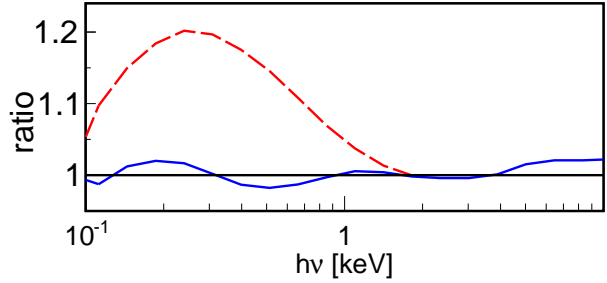}}
\caption{Residuals to fits with a power-law (dashed, red) and a power-law + DISKBB (solid, blue) model of the ADAF MC spectrum for model s2.
The DISKBB spectrum is for $kT= 0.17$ keV and in both cases the power-law index is $\Gamma = 1.7$.
}
\label{fig:ratio} 
\end{figure}	

In models with $\lambda_{2-10} \sim (1-5) \times 10^{-4}$, typically only a weak or moderate first scattering bump occurs below a few keV and at higher energies an approximate power-law shape is formed. We point out that the excess above the power-law extrapolated down to the soft X-ray range can be misinterpreted as being due to the blackbody emission with the temperature, $kT_{\rm BB}$, of several hundred eV. This is illustrated in Fig.\ \ref{fig:ratio}, where the ratio of the ADAF spectrum to a power-law spectrum  shows a pronounced excess below 1 keV, resembling a thermal-like component, whereas the ratio to the same power-law with an additional DISKBB (Mitsuda et al.\ 1984) component shows deviations not exceeding 2 per cent. The relative amplitude of such an artificial thermal-like component accounting for the scattering bump, as measured by the ratio of its and the power-law fluxes in the 0.1 -- 1 keV range, $N_{\rm BB}$, is typically  $N_{\rm BB} \simeq 0.1$--0.5.

The above property is crucial for assessments of the extent of the cold disc at small $\lambda_{2-10}$. E.g.\ Reis et al.\ (2009) claim that an excess below 2 keV in the low-luminosity spectrum of XTE J1118+480 represents the evidence of a thermal disc component with $kT_{\rm BB} \simeq 0.2$ keV, which would imply a disc extending close to the black hole, contrary to the estimates by Esin et al.\ (2001). However, their study relies on the assumption that a strictly power-law spectral component extends between $\sim 100$ eV and 100 keV. We emphasize that a proper physical description of Comptonized-radiation spectra 
is particularly important at low luminosities in assessments of the presence of additional spectral components.

\subsubsection{Turning point}
\label{sect:turn}

The change of the sign of the $\Gamma$--$\lambda_{2-10}$ correlation, around $\lambda_{\rm turn} \simeq 2 \times 10^{-3}$,
can be explained by a change of the character of the dominant seed photons for Comptonization, as pointed out by Sobolewska 
et al.\ (2011). In particular, the change may result from the decrease of the cold disc radius down to $\la 50 R_{\rm g}$, leading to a significant irradiation of the flow (see also fig.\ 12 in Veledina et al.\ 2011 and fig.\ 2 in Gardner \& Done 2013). However, there is a caveat to such a scenario, concerning the assessment by Plant et al.\ (2013) of a much larger $r_{\rm tr} \sim 100$ (implying a negligible irradiation) at $\lambda_{2-10}$ an order of magnitude above  $\lambda_{\rm turn}$. Furthermore, the data set (2) shows the change of the correlation at a similar $\lambda_{2-10}$ as other data sets (note that in all cases $\lambda_{2-10}$ represents only the luminosity of the power-law component) although it corresponds to a much stronger irradiation from a cold disc. This indicates that the change may be rather related  with processes internal to hot flows.

In the context of the observed correlation reversal, it is interesting to note that around $\lambda_{\rm turn}$ the hot flow solutions enter the regime where 
Coulomb cooling of ions becomes substantial, which hints that the change of the $\Gamma$--$\lambda_{2-10}$ relation may be related with the change of the flow structure. At $\lambda_{2-10} < \lambda_{\rm turn}$, where the Coulomb cooling of ions is negligible, the change of electron cooling with $\dot m$  is determined mostly by the $\tau \propto \dot m$ scaling, which yields a roughly linear relation between $\Gamma$ and $\log (\lambda)$. At $\lambda_{2-10} \ga \lambda_{\rm turn}$, the Coulomb cooling of ions should result in a decrease of the height scale, leading to an enhanced efficiency of Compton cooling (see Section \ref{sect:beta}). Malzac \& Belmont (2009) and Droulans et al.\ (2010) estimate in luminous hard states with $\tau > 1$, for the case of electron heating dominated by Coulomb interactions,  ion temperatures an order of magnitude lower than predicted by standard ADAF models, which is consistent with the efficient cooling of ions in this regime. 

We have not performed yet calculations for that range of parameters, which could verify whether the related decrease of $H/R$ is sufficient for the reversal of the $\Gamma$--$\lambda_{2-10}$ relation, or if it should be completed by including some additional source of seed photons. Concerning the latter, the efficient cooling of ions may help in  condensation into either  cold clumps  embedded in the flow (considered already by Shapiro, Lightman \& Eardley et al.\ 1976) or an inner cold disk (Qiao \& Liu 2013). The latter scenario is more feasible in large-$\delta$ models, in which we find maximum values of $\Lambda_{\rm ie}/Q_{\rm i}^+$ at $r \sim 20$ (see Section \ref{sect:delta}).

Finally, hard states with $L$ over an order of magnitude above the reversal were observed  both in  XTE J1550-564 and GX 339-4 (e.g.\ Sobczak et al.\ 2000, Zdziarski et al.\ 2004), which is consistent with the increase of $L$ expected in the regime of the efficient Coulomb transfer of energy in models with small $\delta$ (see Section \ref{sect:mdot}). Furthermore, the maximum luminosities of $\sim 0.2 L_{\rm Edd}$ favour large values of $a$  in both objects (see Section \ref{sect:spin}).

\subsection{Low/moderate-luminosity AGNs}

\subsubsection{$\Gamma$--$\lambda_{2-10}$ plane}

In Fig.\ \ref{fig:observ}bc we compare the $\Gamma$--$\lambda_{2-10}$ relation predicted in our model for $M =2 \times 10^8\, \msun $ with the observational data for low-luminosity AGNs. The model points shown in Fig.\ \ref{fig:observ}bc represent changes of all crucial parameters, as described in Section \ref{sect:3}. As we can see, for a broad range parameters the model predicts a rather narrow distribution of $\Gamma$ at a given $\lambda_{2-10}$. Fig.\ \ref{fig:observ}b shows that there seems to be no systematic agreement between the model and the data for local Seyfert galaxies used by Gu \& Cao (2009), who originally established the X-ray hardening with increasing luminosity. Apparently,  AGNs have a much larger spread of $\Gamma$.

However, results of the X-ray spectral fitting adopted by Gu \& Cao (2009) rely on a rather simplified approach to spectral modelling (cf.\ Cappi et al.\ 2006),  which allows to estimate average trends, however, the assumed models are too simple  for a precise comparisons with the model predictions. Sources whose spectra are measured with high-quality statistics typically require complex models of the continuum and absorption and/or reflection features. A further shortcoming of the data sample presented in Fig.\ \ref{fig:observ}b is  related to the  coverage of rather narrow energy band by {\it XMM} or {\it Chandra}, which is not
sufficient to precisely disentangle the intrinsic X-ray component from emission/absorption.

To reduce the above limitations, we gathered from literature the values of the intrinsic X-ray slopes from detailed models of broad-band data sets (including {\it Integral}, I, and/or {\it Suzaku}, S, data) of several well studied, nearby objects:  NGC 4258 (S, Yamada et al.\ 2009), Circinus (S, Yang et al.\ 2009), Centaurus A (I, Beckmann et al 2011), NGC 4151 (I,S,{\it XMM}, dim and bright states; Lubi\'nski et al 2010) and NGC 5548 (S, Brenneman et al 2012). Crucially for precision of $\lambda_{2-10}$, the first three have $M$ estimated directly from gas or maser kinematics; for Circinus we use  $M=(1.7 \pm 0.3) \times 10^6 M_\odot$ (Greenhill et al.\ 2003), for Cen A, $M = 9.6^{+2.5}_{-1.8} \times 10^7 M_\odot$ (Gnerucci  et al.\ 2011), and for NGC 4258, $M = 3.6 \times 10^7 \msun$ (Miyoshi et al.\ 1995; uncertainty not reported). For the remaining two, $M$ is found from reverberation mapping, $M=6.7 (\pm 2.6)  \times 10^7 M_\odot$ for NGC 5548 (Peterson et al.\ 2004)  and $M=4.6 (\pm 0.6) \times 10^7 M_\odot$ for NGC 4151 (Bentz et al.\ 2006);  the estimated errors do not take into account much larger systematic uncertainties. Finally, we use the fit to the X-ray flux -- photon index anticorrelation in NGC 7213 from Emmanoulopoulos et al.\ (2012) and to find  $\lambda_{2-10}$ we use $M = 8 \times 10^7 M_\odot$ (Schnorr-M\"uller et al.\ 2014) inferred from velocity dispersion and subject to large  uncertainty; we plot confidence limits on the fitted trend, however, we do not plot error bars on $\lambda_{2-10}$ (the upper limit of $M = 2.4 \times 10^8 M_\odot$ from Schnorr-M\"uller et al.\ 2014 would make it close to NGC 4258). 

As we see, objects in Fig.\ \ref{fig:observ}c show a much tighter correlation between $\Gamma$ and $\lambda_{2-10}$. Interestingly, it roughly agrees with the relation, $\Gamma = -0.09 \log \lambda_{2-10} + 1.42$, obtained by Gu \& Cao (2009); a flat, with $\Gamma < 1.7$, spectrum of the intrinsic X-ray emission of the (obscured) Seyfert nucleus in Circinus indicates that it extends  up to  $\lambda_{2-10} \simeq 0.005$. We see that the standard ADAF  model predicts spectra systematically harder than those of the well studied AGNs, and discrepancies increase with increasing $\lambda_{2-10}$. The two model points located close to Cen A are for the models (a8 and o3) with very tenuous flows, with $\tau^{\rm PS} \sim 0.01$ and $kT_{\rm e}^{\rm PS} \sim 1$ MeV, and their X-ray spectra strongly deviate from a power-law.

\subsubsection{Electron temperature}

The disagreement with the observed $\Gamma$--$\lambda_{2-10}$ indicates that the thermal synchrotron radiation provides an insufficient input of seed photons 
in hot flows powering AGNs. This conclusion is further confirmed by the comparison with electron temperatures estimated from the cut-off energy. Such measurements are available only for a few X-ray--brightest AGNs, furthermore, there is a discrepancy in temperatures estimated from data provided by different detectors (e.g., see Petrucci et al.\ 2001 and Lubi\'nski et al.\ 2010). For our comparison we adopt results from studies used to plot   Fig.\ \ref{fig:observ}c.
Fig.\ \ref{fig:comppsparam}c shows parameters of the COMPPS fits for NGC 4151 (Lubi\'nski et al.\ 2010; slab fits), Cen A (Beckmann et al.\ 2011; slab) and
NGC 5548 (Brenneman et al.\ 2012; sphere fit so we plot a reduced $\tau$). We can see a clear disagreement with the plasma parameters predicted by the model. 
An apparently small disagreement of the model for $\beta=0.3$, $\delta=10^{-3}$  and $\dot m=0.1$ with  some AGNs  turns out to be substantial when the dependence on luminosity is taken into account; this model is indicated by triangles in Fig.\ \ref{fig:observ}c.

Interestingly, in the ($L/\ledd$)--$\tau^{\rm PS}$ plane (Fig.\ \ref{fig:taulum}) the bright state of NGC 4151 is slightly above the ($\beta=0.3$, $\delta=10^{-3}$) branch, while   Cen A and dim NGC 4151 lie on the high-$\beta$  branch. NGC 5548 is close to high-$\delta$ solutions, which is particularly curious because it has a similar Eddington ratio to bright NGC 4151, however, its $M$ is rather poorly constrained and it may be also close to the ($\beta=9$, $\delta=10^{-3}$) branch. In principle, the above differences may indicate the change of physical parameters of hot flows in AGNs, e.g.\ the increase of $\beta$ with decreasing $\dot m$. We note, however, that these observational constraints on the optical depth strongly depend on the quality of hard X-ray data (essential for reliable determination of $T_{\rm e}$ in spectral fits), which is typically poor in AGNs.  Note also that for AGNs the total luminosity in Fig.\ \ref{fig:taulum} is determined as $7 L_{2-10}$ (which scaling factor is found in our model spectra with relevant $\Gamma$ and $T_{\rm e}^{\rm PS}$). Assessments of bolometric luminosities of these AGNs, available in literature, are not suitable for the comparison with our hot flow model, as they typically take into account other spectral components, e.g.\ radiation reprocessed in optically thick matter.

\subsubsection{Extent of a cold disc}

Consistently with the hot flow scenario, and indicating a negligible irradiation of the central region by a cold disc, neither of the sources in Fig.\ \ref{fig:observ}c shows  signatures of a relativistically distorted reflection component. For example, the width of the Fe K$\alpha$ line (a few tens of eV)  in high-quality {\it XMM} or {\it Suzaku} data of NGC 4151 and NGC 5548 (Lubi\'nski et al 2010, Brenneman et al 2012) implies the lack of an optically thick
material within at least the innermost $\sim 100 R_{\rm g}$.

Studies of the optical/UV emission, presumably representing the thermal emission of a cold disc, give consistent hints for large values of $r_{\rm tr}$; e.g., for NGC 5548, $r_{\rm tr} \simeq 100$--200 is estimated in modelling the broad-band spectrum (Chiang \& Blaes 2003) and the variability properties (Czerny et al.\ 1999).
In turn, a much smaller  $r_{\rm tr} \simeq 15$ is assessed by Lubi\'nski et al.\ (2010) as giving an approximate equipartition between the optical/UV and the X-ray emissions, which is observed in NGC 4151. We note that such (or even smaller for large $a$) $r_{\rm tr}$ gives equal luminosities of an inner flow (at $r < r_{\rm tr}$) and outer disc (at $r > r_{\rm tr}$) only for large $\delta$ ($\sim 0.5$). For small $\delta$, such an equipartition corresponds to $r_{\rm tr} \simeq 300$, more consistent with the observed narrowness of the Fe K$\alpha$ line.

\section{Summary and discussion}
\label{sect:summary}

We have thoroughly  studied spectral properties of radiation produced by thermal Comptonization of thermal synchrotron radiation in hot flows.
Two major trends in the dependence on the model parameters are related with (1) the magnitude of electron heating, which depends primarily on the amount of the dissipated energy which is used for the direct heating of electrons; and (2) the geometrical thickness of the flow, which depends primarily on the magnetic field strength and, for weak magnetic fields, on the presence of an outflow. Large changes of parameters determining these properties, $\beta$, $\delta$ and (only for large $\beta$) $s$, lead to the change of $E_{\rm cut}$ by hundreds of keV. On the other hand, the  $\Gamma$--$\lambda_{2-10}$ relation is not strongly affected by their changes, although large-$\delta$ models predict systematically softer spectra (by up to $\Delta \Gamma \simeq 0.2$) at a given $\lambda_{2-10}$.

The black-hole spin parameter determines the radiative efficiency of flows with large $\delta$ and, then, there is some degeneracy between the values of $\delta$ and $a$. Apart from this, $a$ has a smaller effect on spectral shape than $\delta$ or $\beta$. We do not find a significant dependence of spectra on the viscosity parameter; in models with small $\delta$, there is a degeneracy between $\alpha$ and $\dot m$. Also the outflow does not affect spectral properties significantly, besides lowering $T_{\rm e}$ in models with large $\beta$.

Basing on the ion heating-to-cooling ratio, which determines $\dot m$ at which a local transition to a cold disc should occur, we expect that larger maximum luminosities can be achieved for small $\delta$, large $\alpha$ and large $a$. Such values of these parameters are favoured to explain large luminosity, above $0.1 \ledd$, observed in some hard states, especially considering a large value of $r_{\rm tr} \sim 100$ estimated in GX 339-4 at such large $L$ by Plant et al.\ (2013).

We find that the $\Gamma$--$\lambda_{2-10}$ relation predicted by our model agrees with that observed at $L \la 0.01 \ledd$ in well studied black-hole transients,
except for the declines of strong outbursts where a weakly-recessed disc contributes significantly down to $L \sim 10^{-3} \ledd$. Then, the observed $\Gamma$--$\lambda_{2-10}$ requires a weak input of seed photons, consistent with the thermal synchrotron origin, which puts some caveats on the general framework of hybrid models (see below).
It also poses a major challenge for the hot flow model, making it difficult to reconcile with a low $kT_{\rm e} \sim 100$ keV (which would require a stronger cooling) at $L \sim 10^{-3} \ledd$, such as found in {\it BeppoSAX} observations of XTE J1118+480 by Frontera et al.\ (2003); as we note in Section \ref{sect:te}, however, this measurement may be affected by discrepancies between phenomenological Comptonization and the actual hot-flow spectra below $E_{\rm cut}$.

Note that we considered  angle-averaged emission of hot flows. This should not affect our comparison with the transient systems. Only face-on, with  $\theta_{\rm obs} < 30$ deg, model spectra  show apparent luminosities significantly lower from the angle-averaged ones, whereas the systems considered here have rather high orbital inclination, with $\ga 70$ deg in both XTE J1550-564 (Orosz et al.\ 2011) and XTE J1118+480 (Khargharia et al.\ 2013),  and a low inclination of GX 339-4 disfavoured by Zdziarski et al.\ (2004).

We emphasize that at $\lambda_{2-10} \la 10^{-4}$ the Comptonized spectra deviate significantly from the power-law shape in the soft X-ray range, which effect is relevant for investigation of other spectral components, in particular thermal emission of a cold matter. At still lower luminosities, $\lambda_{2-10} \la 10^{-5}$, Plotkin et al.\ (2013) find a constant spectral index, $\Gamma \simeq 2.1$, over at least 4 orders of magnitudes in $\lambda_{2-10}$. This range of luminosities is beyond the applicability of our model, however, we note hints for flattening of the $\Gamma$--$\lambda_{2-10}$ relation in   small-$\dot m$   models with heating of electrons dominated by compression (i.e.\ those with small  $\delta$).

Discussions of the luminous hard-state are often based on observations of the persistent black hole binary, Cyg X-1. It exhibits the known luminous spectral
states observed in transient black-hole binaries, which indicates that similar  physical mechanisms are involved. However, it does not enter the quiescent or the low-luminosity hard state, in which the  disappearance of an inner cold disc occurs in transient objects. Then, we may suspect that irradiation of the hot flow by the  thermal disc emission is always significant and hence the observational data for  Cyg X-1 do not provide useful constraints on the plain hot-flow emission model. In the simplest scenario, motivated by the behaviour of transient systems, we may expect that the value of $r_{\rm tr}$ corresponds to the minimum luminosity reached by the object since the last preceding soft state, so the magnitude of irradiation by the cold disc may be different for various observations at a given $L$. In agreement with such a complex dependence on the source history, Zdziarski et al.\ (2011) find that in the hard state the spectral index changes within $\simeq 1.4-1.9$ with $L$ changing by a factor of $\sim 4$, but with no systematic correlation and also the trend of $T_{\rm e}$ with $L$ is uncertain. 

Finally, we note a considerable dependence on the black hole mass, with models involving supermassive black holes characterised by significantly higher electron temperatures, and harder X-ray spectra, than these with stellar-mass black holes. This property is clearly inconsistent with observations. In contrary to black hole binaries, the best-studied AGNs, observed at $L \simeq (0.001$--$0.01) \ledd$, disagree with predictions of the purely thermal version of the model. As these AGNs show strong hints for the lack of an optically thick disc within the innermost several hundred $R_{\rm g}$,  the disagreement indicates the need for a revision of the model for hot flows around supermassive black holes.

A relevant extension of the model likely concerns a nonthermal electron component, such as considered in hybrid models. Indeed, the hybrid model  may successfully explain the soft spectra and low values of $T_{\rm e}$ in AGNs (see Veledina et al.\ 2011); similar, preliminary results for the hot-flow model implementing nonthermal electrons from the decay of charged pions are presented in Nied\'zwiecki et al.\ (2014).

The hybrid model could also provide an attractive explanation to the too-high--$T_{\rm e}$ problem in black hole binaries. However, as noted above, presence of a significant nonthermal component in black-hole transients at $L \la 0.01 L_{\rm Edd}$ would lead to disagreement with the observed $\Gamma$--$\lambda_{2-10}$.
Observationally, the presence of nonthermal electrons is mainly motivated by detections of nonthermal tails (e.g.\ McConnell et al.\ 2002, Droulans et al.\ 2010) which usually involve observations at $L \ga 0.01 L_{\rm Edd}$, although there are also reports of high-energy tails at lower $L$ (e.g.\ Bouchet et al.\ 2009).
In the hard state, the tails may be related with either a non-fully recessed disc (as suggested by the larger strength of the tail observed in the soft state in Cyg X-1) or jet (Zdziarski et al.\ 2014), or it may indicate that a strong nonthermal component indeed appears in hot flows at large $L$. Our model including the nonthermal component in hot flows is currently under development. 

We also intend to investigate self-consistent solutions for high $L$, corresponding to LHAF regime, which is particularly important as most of the precise measurements of $E_{\rm cut}$ are  available for such high $L$. It will also clarify if strong Coulomb cooling of ions may be a relevant effect for explaining 
the observed reversal of the $\Gamma$--$\lambda_{2-10}$ correlation sign. The most detailed study of LHAF solutions is presented in Xie \& Yuan (2012), however, it uses a local Compton as well as a non-relativistic model and we note that both approximations are risky in that regime. For example, the same set of parameters may give an LHAF solution in the local Compton and no self-consistent solution in the global Compton  model (see the model with $\dot m_{\rm out}=1$ in Xie et al.\ 2010). We may also expect that GR LHAFs, for which dissipative heating of ions should dominate for large $a$,  have different properties than non-relativistic LHAFs, for which compressive heating of ions strongly dominates.

\section*{ACKNOWLEDGMENTS}

We thank P.~Lubi\'nski for useful discussion and the referee for useful suggestions. 
This research has been supported in part by the Polish NCN grant N N203 582240. FGX is supported by the National Basic Research Program of China (973 Program, grant 2014CB845800), the NSFC (grants 11203057, 11103061, 11133005 and 11121062), and the Strategic Priority Research Program "The Emergence of Cosmological Structures" of the Chinese Academy of Sciences (Grant XDB09000000).

\label{lastpage} 

\end{document}